# *Spontaneous Breaking of Time Reversal Symmetry and Time-Crystal States in Chiral Atomic Systems*

*Mário G. Silveirinha*[(1)*], *Hugo Terças*[(2)], *Mauro Antezza*[(3,4)]

[(1)] *University of Lisbon–Instituto Superior Técnico and Instituto de Telecomunicações, Avenida Rovisco Pais, 1, 1049-001 Lisboa, Portugal*

[(2)] *GoLP/Instituto de Plasmas e Fusão Nuclear, Instituto Superior Técnico, 1049-001 Lisboa, Portugal*

[(3)] *Laboratoire Charles Coulomb (L2C), UMR 5221 CNRS-Université de Montpellier, F-34095 Montpellier, France*

[(4)] *Institut Universitaire de France, 1 rue Descartes, F-75231 Paris Cedex 05, France*

**Abstract**

We present a theoretical study of the interaction between an atom characterized by a degenerate ground state and a reciprocal environment, such as a semiconductor nanoparticle, without the presence of external bias. Our analysis reveals that the combined influence of the electron's intrinsic spin magnetic moment on the environment and the chiral atomic dipolar transitions may lead to either the spontaneous breaking of time-reversal symmetry or the emergence of time-crystal-like states with remarkably long relaxation times. The different behavior is ruled by the handedness of the precession motion of the atom's spin vector, which is induced by virtual chiral-dipolar transitions. Specifically, when the relative orientation of the precession angular velocity and the electron spin vector is as in a spinning top, the system manifests time-crystal-like states. Conversely, with the opposite relative orientation, the system experiences spontaneous symmetry breaking of time-reversal symmetry. Our findings introduce a novel mechanism for the spontaneous breaking of time-reversal symmetry in atomic systems, and unveil an exciting opportunity to engineer a nonreciprocal response at the nanoscale, exclusively driven by the quantum vacuum fluctuations.

---

[*] To whom correspondence should be addressed: E-mail: *mario.silveirinha@tecnico.ulisboa.pt*

# I.Introduction

Time crystals refer to a phase of matter where the ground state of a system described by a time-independent Hamiltonian exhibits repetitive oscillations in time without external driving [1-6]. This concept was introduced by Wilczek and Shapere [1-2] and is rooted in a spontaneous symmetry breaking of the time-translation symmetry of the system Hamiltonian. Spontaneous symmetry breaking is a fundamental mechanism observed in various physical processes such as ferroelectricity, ferromagnetism, and superconductivity [7, 8, 9].

At the microscopic level, physical systems are typically governed by time-reversal invariant Hamiltonians [10]. This means that if a specific time dynamics complies with the laws of physics, its time-reversed counterpart is also valid and can be observed in a real physical experiment if the system can be prepared in the required initial state. For atomic systems with half-integer total spin (including nuclear spin), the time reversal symmetry dictates that the eigenstates of the system are organized in doublets, which are pairs of degenerate states with the same energy. This result is known as Kramers' theorem [11], and it implies that the ground state of such systems is necessarily doubly degenerate.

This work aims to demonstrate theoretically that the interaction between an atomic system with a degenerate ground state and a reciprocal environment (e.g., a metallic or semiconductor nanoparticle without any form of external bias) can lead to either the spontaneous breaking of time-reversal symmetry or the formation of time-crystal-like states with extraordinarily long relaxation times. Our analysis reveals that the key to unlocking such extreme physics lies in the combination of chiral-type atomic transitions and the influence of the electron's intrinsic spin magnetic moment on the environment. In particular, we show that chiral-dipolar transitions may cause the spin magnetic moment of the electron to undergo a periodic precession motion around the symmetry axis defined by the atom and nanoparticle ($z$-axis). We demonstrate that when the relative orientation of the precession angular velocity ($\boldsymbol{\omega}_\mathrm{m}$) and the electron spin vector ($\hat{\mathbf{S}}$) is the same as for a spinning top ($\boldsymbol{\omega}_\mathrm{m} \sim \left(\hat{\mathbf{S}} \cdot \hat{\mathbf{z}}\right)\hat{\mathbf{z}}$), the orbit of the spin vector is periodic in time and resilient to perturbations of the system Hamiltonian. In such a case, the atomic system supports time-crystal-like states in its ground state. In contrast, when the precession velocity satisfies $\boldsymbol{\omega}_\mathrm{m} \sim -\left(\hat{\mathbf{S}} \cdot \hat{\mathbf{z}}\right)\hat{\mathbf{z}}$, the system orbit is dragged to a well-defined attractor point in the Bloch sphere resulting in a spontaneous symmetry breaking of the time-reversal symmetry.

It should be noted that quantum "time-crystals" usually refer to quantum many-body systems such that the single-particle expectation of one-body operators is independent of time in the ground state, while the expectation of $N$-body operators can be time dependent [5]. Here, we employ the term "time-crystal" in a broader sense to describe a quantum system capable of undergoing periodic (metastable) motion when it interacts with a dissipative environment. Due to this reason, our findings are not affected the "no-go" theorems previously discussed in the literature [5, 12, 13]. In fact, in agreement with [12] we find that our time-crystal-like states are destroyed by the effects of radiation loss. Within the validity of our model, the time scale of such a process is so large that it can be neglected for all practical purposes.

As mentioned above, we find that for some classes of atomic systems the interaction of the spin magnetic moment of the electron with the environment results in a spontaneous symmetry breaking of the time-reversal symmetry. Material platforms with broken time reversal symmetry are particularly interesting from an electromagnetic point of view as they can lead to nonreciprocal interactions, and enable the unidirectional propagation of light and optical isolation [14-23].

Moreover, over the past decade, there has been a significant surge of interest in nonreciprocal effects in quantum systems. In general, nonreciprocity in the quantum realm can be achieved through various means, such as applying an external magnetic bias [24-27], preparing the system in a chiral quantum state [24, 27-31], utilizing non-linearities [32], or employing a nonreciprocal substrate [33-35]. When a quantum system is prepared in a state that favors specific circularly polarized (chiral) optical transitions, it exhibits remarkable directional properties [24, 36-37]. The nonreciprocity of the environment can uniquely influence the Casimir-Polder dispersion forces and torques in atomic and nano-scale systems [34-35, 38-41]. Additionally, driven time-variant systems provide peculiar environments for exotic quantum light-matter interactions [42-43].

Unlike previous studies, our work presents a mechanism for the spontaneous breaking of time reversal symmetry in an atomic system, solely based on the interactions between the atom and the quantum vacuum. No external bias is required beyond the inherent vacuum fluctuations. This unique mechanism leads to the emergence of a nonreciprocal gyrotropic response in the atomic system. This effect bears resemblance to the phenomena of ferromagnetism and anti-ferromagnetism observed in solid-state materials. The spontaneous symmetry breaking can only occur in the presence of chiral transitions, which may be attributed to the influence of the atomic spin-orbit interaction.

The article is organized as follows. In Sect. II, we introduce the geometry of the problem and the atomic model of interest. In Sect. III, we derive a reduced nonlinear master equation that describes the time-evolution of the atomic degrees of freedom. In Sect. IV, we demonstrate that, for the ground-subspace, the master equation can be written in the von Neumann form using a nonlinear effective Hamiltonian. By employing a quasi-static approximation, we obtain an explicit analytical formula for the effective Hamiltonian. In Sect. V, we show that the system can exhibit time-crystal-like states or broken time-reversal symmetry, depending on the handedness of the chiral dipolar transitions with respect to the electron spin. Finally, Sect. VI provides a brief summary of the main results.

## II. The Kramers qubit

The geometry of the problem is represented in Fig. 1. It consists of a time-reversal invariant atomic system interacting with a spherical plasmonic nanoparticle of radius $R$. The distance between the atom and the nanoparticle center is $d$. Due to the Kramers theorem, all the atomic states must be degenerate [11]. Here, we consider a minimal model for such a system corresponding to a two-level atom formed by two degenerate ground states ($|g_i\rangle$, $i$=1,2) and two or more degenerate excited states ($|e_j\rangle$, $j$=1,2,...). For simplicity, in the main text it is supposed that the excited states subspace has dimension 2, but the theory can be readily generalized to an excited state subspace with higher dimension (see Appendices A and B). Following [44], we refer to such a system as a "Kramers two-level system" or "Kramers qubit".

The atomic states are connected by the time-reversal operator $\mathcal{T}$ as $|g_2\rangle = \mathcal{T}|g_1\rangle$ and $|e_2\rangle = \mathcal{T}|e_1\rangle$. As $\mathcal{T}^2 = -\mathbf{1}$, the Kramers pairs also satisfy $|g_1\rangle = -\mathcal{T}|g_2\rangle$ and $|e_2\rangle = -\mathcal{T}|e_1\rangle$ [11, 44]. The electric dipole moment operator is $\hat{\mathbf{p}}_e = \hat{\mathbf{p}}_e^- + \hat{\mathbf{p}}_e^+$, with $\hat{\mathbf{p}}_e^+ = (\hat{\mathbf{p}}_e^-)^\dagger$ and $\hat{\mathbf{p}}_e^- = \gamma_d \sigma_{g1,e1} + \gamma_d^* \sigma_{g2,e2} + \gamma_c \sigma_{g1,e2} - \gamma_c^* \sigma_{g2,e1}$, with $\sigma_{g_i,e_j} = |g_i\rangle\langle e_j|$ [44]. The four transition electric dipole moments $(\gamma_d, \gamma_d^*, \gamma_c, \gamma_c^*)$ between the excited and ground states are represented in the top panel of Fig. 1. As discussed in Ref. [44], the presence of spin-orbit coupling generally prevents the selection of a basis for the ground and excited states without crossed transitions. In general, the spin-orbit interaction implies that $\gamma_c \neq 0$ and $\gamma_d \neq 0$.

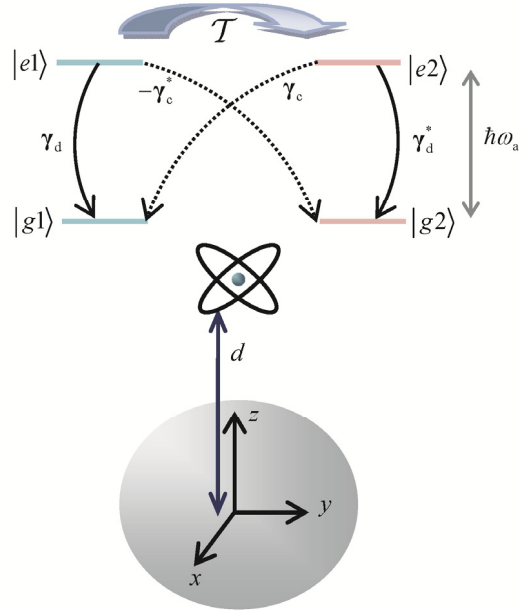

**Fig. 1** A Kramers two-level system stands at a distance $d$ from a plasmonic nanoparticle. The energy levels and the transition dipole-moments of the two-level system are shown in the top panel. The two ground and the two excited states are related by the time-reversal operator $\mathcal{T}$.

Recently, it was suggested that the Jahn-Teller trigonal molecular systems $X_3$ (e.g., the alkali trimers $Li_3$, $Na_3$ and $K_3$) may be good candidates to implement a Kramers two-level system [45]. It has been predicted that when such molecular systems are placed nearby a planar substrate the global rotational symmetry may be spontaneously broken by the quantum vacuum fluctuations [45], in agreement with the theory of Ref. [44]. In the following, we further motivate the Kramers two-level model by showing that it emerges naturally from the Schrödinger equation with the spin-orbit interaction.

*A. Hydrogen-like model for the Kramers two-level system*

Let us consider the Schrödinger equation for a hydrogen-like atom described by a spherical symmetric electric potential including the spin-orbit interaction [46]. The effects of nuclear spin are disregarded in the analysis. It is implicit that the nuclear spin is an integer to ensure that the total spin is half-integer (e.g., deuterium, which is an isotope of hydrogen, has such a property).

The spin-orbit interaction intertwines the spin and orbital degrees of freedom, in such a way that it becomes impossible to write the energy eigenfunctions as a product of some coordinate dependent function (the orbital part) and a coordinate independent spinor (the spin part). For example, the first few energy levels of a hydrogen-like atom are the $1S$, $2S$, $2P$ states. The spin orbit interaction splits the $2P$ states into two levels $2P^{1/2}$ and $2P^{3/2}$ [46]. The superscript $j$ determines the total angular momentum. The $2S$ states are excluded from additional considerations because by symmetry they cannot be coupled to the $1S$ ground states, i.e., the corresponding transition electric dipole moment vanishes. The energy of the $2P^{1/2}$ states is lower than the energy of the $2P^{3/2}$ states. So, the light interactions with a hypothetical hydrogen-like atom with strong spin orbit coupling can be approximately described by dipolar transitions between the $1S^{1/2}$ (ground states) and the $2P^{1/2}$ states (excited states). Both the $1S^{1/2}$ and $2P^{1/2}$ states are doublets. The excited states can be written explicitly as:

$$|e_1\rangle \equiv |j=1/2, m_j=-1/2\rangle = \frac{1}{\sqrt{3}}|l=1, m_l=0,\downarrow\rangle - \sqrt{\frac{2}{3}}|l=1, m_l=-1,\uparrow\rangle. \quad (1a)$$

$$|e_2\rangle \equiv |j=1/2, m_j=1/2\rangle = \frac{1}{\sqrt{3}}|l=1, m_l=0,\uparrow\rangle - \sqrt{\frac{2}{3}}|l=1, m_l=1,\downarrow\rangle. \quad (1b)$$

The labels $(j, m_j)$ refer to the total angular momentum and the labels $(l, m_l)$ to the orbital angular momentum. The arrows indicate the states spin up or spin down with respect to $z$. Note that the spin state cannot be factored out. It can be easily checked that the two excited states are linked as $|e_2\rangle = \mathcal{T}|e_1\rangle$. The two ground states are simply given by:

$$|g_1\rangle \equiv |l=0,\uparrow\rangle, \qquad |g_2\rangle \equiv |l=0,\downarrow\rangle. \quad (2)$$

It is implicit that the principal quantum number of the ground and excited states is $n=1$ and $n=2$, respectively. Evidently, the transition dipole moments are given by $\boldsymbol{\gamma}_d = \langle g_1|\hat{\mathbf{p}}_e|e_1\rangle$ and $\boldsymbol{\gamma}_c = \langle g_1|\hat{\mathbf{p}}_e|e_2\rangle$ with $\hat{\mathbf{p}}_e = q\hat{\mathbf{r}}$ the electric dipole operator. Here, $q=-e$ is the negative charge of the electron. Taking into account the spatial symmetry of the $S$ and $P$ states it can be easily shown that the two transition dipole moments assume the form:

$$\boldsymbol{\gamma}_d = \frac{\gamma_d}{\sqrt{2}}(\hat{\mathbf{x}} - i\hat{\mathbf{y}}), \qquad \boldsymbol{\gamma}_c = \gamma_c \hat{\mathbf{z}}. \quad (3)$$

Therefore, one of the dipolar transitions is circularly polarized (chiral), whereas the other dipolar transition is linearly polarized. In the hydrogen-like model the amplitudes of the two dipoles are related as $|\gamma_d| = \sqrt{2}|\gamma_c|$, so that the chiral transitions dominate.

This analysis confirms that even in very elementary systems the spin-orbit interaction originates crossed dipolar transitions that couple the spin up and spin down ground states. Evidently, the considered toy model describes an isotropic atomic system, i.e., invariant under arbitrary space rotations. As detailed later, in our analysis we shall consider more general systems with reduced symmetry ($|\gamma_d| \neq \sqrt{2}|\gamma_c|$). For example, following Ref. [45], the Jahn-Teller molecular systems $X_3$ are characterized by a similar Kramers model but with $\gamma_c \approx 0$. The $X_3$ molecules are planar and have a 3-fold rotation symmetry.

### B. Magnetic dipole due to the intrinsic electron spin

Besides the electric dipole transitions, in our model we also consider the magnetic dipole coupling that arises due to the intrinsic spin magnetic moment of the electron. The spin magnetic moment operator $\hat{\mathbf{m}}_s$ is written in terms of Pauli matrices in the usual way.

The interesting point is that the intrinsic spin magnetic moment may create a static magnetic field $\mathbf{B}_0$ which can tailor the response of the plasmonic nanoparticle. The static magnetic field can be found using the dipole approximation [47]

$$\mathbf{B}_0 = \frac{\mu_0}{4\pi d^3}(3\hat{\mathbf{z}} \otimes \hat{\mathbf{z}} - \mathbf{1}) \cdot \mathbf{m}_s, \qquad (4)$$

with $\mathbf{m}_s = \langle \hat{\mathbf{m}}_s \rangle$ the expectation of $\hat{\mathbf{m}}_s$ and $d \geq R$ the distance between the atom and the nanoparticle. For simplicity, we disregard any orbital contributions to $\mathbf{B}_0$, as they are much weaker.

Importantly, the field $\mathbf{B}_0$ can influence the macroscopic response of a free-electron gas. Specifically, under a static magnetic bias ($\mathbf{B}_0 = B_0 \hat{\mathbf{u}}$) a Drude plasma acquires a gyroelectric response determined by the permittivity tensor [48, 49]:

$$\frac{\overline{\varepsilon}}{\varepsilon_0} = \varepsilon_t \mathbf{1}_t + i\varepsilon_g \hat{\mathbf{u}} \times \mathbf{1} + \varepsilon_a \hat{\mathbf{u}} \otimes \hat{\mathbf{u}}, \tag{5}$$

with $\mathbf{1}_t = \mathbf{1} - \hat{\mathbf{u}} \otimes \hat{\mathbf{u}}$. The permittivity elements are:

$$\varepsilon_t = 1 - \frac{\omega_p^2 (1 + i\Gamma/\omega)}{(\omega + i\Gamma)^2 - \omega_0^2}, \qquad \varepsilon_g = \frac{1}{\omega} \frac{\omega_p^2 \omega_0}{\omega_0^2 - (\omega + i\Gamma)^2}, \qquad \varepsilon_a = 1 - \frac{\omega_p^2}{\omega(\omega + i\Gamma)}. \tag{6}$$

Here, $\omega_0 = -qB_0/m^*$ is the cyclotron frequency determined by the bias magnetic field $\mathbf{B}_0 = B_0 \hat{\mathbf{u}}$, $\Gamma$ is the collision frequency, $m^*$ is the electron effective mass, and $\omega_p$ is the plasma frequency. Thus, the atomic state controls the response of the nanoparticle, and hence indirectly tailors the environment with which it interacts.

It is relevant to point out that the gyrotropic response of the plasma is itself rooted in a nonlinear response of the material. Specifically, the motion of the electrons at the microscopic level is ruled by the Lorentz force $q(\mathbf{E} + \mathbf{v} \times \mathbf{B})$, and the magnetic part of the force mixes the particle velocity with the magnetic field, corresponding to a nonlinear interaction. The linear gyrotropic response (6) arises after a suitable linearization of the Lorentz force [22, 48].

Using the estimate $\langle m_s \rangle \sim \frac{\hbar}{2} \frac{q}{m_e}$ with $m_e$ the free-electron mass, one sees that the cyclotron frequency amplitude is on the order of $\omega_{0\perp} \equiv \frac{\mu_0}{8\pi d^3} \frac{\hbar q^2}{m_e m^*} = \alpha_{e,fs} \frac{\hbar^2}{2 m_e m^*} \frac{1}{c} \frac{1}{d^3}$ with $\alpha_{e,fs} \approx 1/137$ the electron fine structure constant. For a distance of $d = 5 nm$ and for a semiconductor with $m^* = 0.001 m_e$ one obtains $\omega_{0\perp} \sim 2\pi \times 0.2\,\text{GHz}$ ($B_0 \sim 7.4\,\mu\text{T}$). This estimate shows that the gyrotropic effects due to the spin magnetic moment can be significant and thereby lead to a strong coupling of the atomic system with the environment.

## III. Nonlinear master equation

In Appendix A, we use the Born-Markov approximation to derive a master equation for the reduced density matrix $\hat{\rho}_S$ that describes the time evolution of the atomic degrees of freedom, $\partial_t \hat{\rho}_S = -i\frac{1}{\hbar}\left[\hat{H}_{at}, \hat{\rho}_S\right] + \mathcal{L}\hat{\rho}_S$ [Eq. (A14)]. Here, $\hat{H}_{at}$ is the time-reversal invariant atomic Hamiltonian and $\mathcal{L}$ is a Lindblad operator written in terms of the dipole atomic operators and of the system Green's function $\overline{\mathcal{G}}$ [Eq. (A15)]. As usual, the Born approximation considers that the full-system density matrix satisfies $\hat{\rho}(t) \approx \hat{\rho}_S(t) \otimes \hat{\rho}_E$ at all time instants, with $\hat{\rho}_S$ and $\hat{\rho}_E$ density matrices that depend on the atomic (environment) degrees of freedom, respectively. Importantly, as the atomic state tailors the environment through the magnetic moment $\mathbf{m}_s = \text{tr}\{\hat{\mathbf{m}}_s \hat{\rho}_S\}$, the Green's function depends indirectly on $\hat{\rho}_S$ through the bias magnetic field. Consequently, the Lindblad operator also depends on $\hat{\rho}_S$ and the reduced master equation is *nonlinear*.

The linearity of the Schrödinger equation, or more generally of a master equation, is an essential feature of the quantum superposition principle. Yet, some many-body condensed matter systems may be *effectively* described by nonlinear Schrödinger equations. Well known examples are the Hartree-Fock equation and the Gross-Pitaevskii equation [50-52]. Similarly, we find that interaction of the atomic system with a gyroelectric nanoparticle may be described by a nonlinear master equation due to the effect of the intrinsic spin magnetic moment. It is important to note that the nonlinearity of the master equation does not contradict the superposition principle. In fact, the reduced master equation is supposed to describe the time evolution of systems for which at initial time $\hat{\rho}(t) \approx \hat{\rho}_S(t) \otimes \hat{\rho}_E(\mathbf{B}_0)$, with $\hat{\rho}_E(\mathbf{B}_0)$ the environment ground-state determined by the atomic spin magnetic moment. The superposition of matrices of the type $\hat{\rho}_{S,i}(t) \otimes \hat{\rho}_E(\mathbf{B}_{0,i})$ ($i$=1,2,…) is not an object with the same structure as the individual matrices. Thereby, the time evolution of a density matrix that is a linear combination of matrices of the type $\hat{\rho}_{S,i}(t) \otimes \hat{\rho}_E(\mathbf{B}_{0,i})$ cannot be studied using the reduced (nonlinear) master equation. In other words, the lack of linearity in the reduced master equation is not in conflict with the superposition principle. From a more intuitive point of view, the nonlinearity can be attributed to the fact that the magnetic field expectation for a superposition of two atomic states is not the superposition of the fields associated with the individual states.

## IV. Ground state physics

### A. Effective Hamiltonian

This article is focused on the ground-state physics. In Appendix B, we show that the ground subspace generated by linear combinations of elements of the type $|g_i\rangle\langle g_j|$ is closed. This implies that if the reduced density matrix $\hat{\rho}_S$ at the initial time ($t=0$) belongs to the ground subspace, it will continue to reside within the ground subspace for $t > 0$. Note that an arbitrary initial state will eventually evolve into an element of the ground subspace due to the usual spontaneous emission decay.

Furthermore, it is proven in Appendix B that the dynamics of the ground subspace is described by a simplified reduced master equation of the type $\partial_t \hat{\rho}_S = -i\frac{1}{\hbar}\left[\hat{H}_{ef}, \hat{\rho}_S\right]$, with $\hat{H}_{ef}$ an effective Hamiltonian that depends on the transition dipole operators and on the system Green's function [Eq. (B5)]. Thus, the ground subspace effective dynamics is modeled by a von Neumann equation. This means if the initial state is a pure state of the type $\hat{\rho}_S(t=0) = |\psi_0\rangle\langle\psi_0|$ with $|\psi_0\rangle$ some element in the atomic ground subspace, then for $t > 0$ the reduced density matrix satisfies $\hat{\rho}_S(t) = |\psi(t)\rangle\langle\psi(t)|$ with $|\psi(t)\rangle = c_1(t)|g_1\rangle + c_2(t)|g_2\rangle$, i.e., the density matrix describes a pure state at all time instants. The time evolution of the coefficients $c_1(t), c_2(t)$ is ruled by a nonlinear Schrödinger equation [Eq. (B8)]

$$i\hbar\partial_t \begin{pmatrix} c_1 \\ c_2 \end{pmatrix} = \mathbf{H}_{ef}(c_1, c_2) \cdot \begin{pmatrix} c_1 \\ c_2 \end{pmatrix}, \tag{7}$$

where $\mathbf{H}_{ef} = \left[h_{m,n}(\mathbf{B}_0)\right]_{m,n=1,2}$ is a 2×2 matrix with entries:

$$h_{m,n}(\mathbf{B}_0) = -\mathrm{tr}\left\{\mathbf{Q}^{(m,n)} \cdot \left(\frac{1}{2\pi}\int_{0^+}^{\infty} d\xi \frac{\overline{\mathcal{G}}(\mathbf{r}_0, \mathbf{r}_0, i\xi)}{\omega_a - i\xi} + \frac{\left[\overline{\mathcal{G}}(\mathbf{r}_0, \mathbf{r}_0, i\xi)\right]^T}{\omega_a + i\xi}\right)\right\}. \tag{8}$$

In the above, $\omega_a$ is the atomic transition frequency of the two-level system, $\mathrm{tr}\{...\}$ is the trace operator, and $\mathbf{Q}^{(m,n)} = \langle g_m | \hat{\mathbf{p}}^- \otimes \hat{\mathbf{p}}^+ | g_n \rangle$ is a 6×6 matrix determined by the atomic dipole operators. Note that the system Green's function depends on $\mathbf{B}_0$, and

consequently the matrix operator $\mathbf{H}_{\text{ef}}$ depends explicitly on $c_1(t), c_2(t)$, leading to a nonlinear dynamics.

## B. Quasi-static approximation

In Appendix C, we evaluate explicitly the elements of the effective Hamiltonian neglecting retardation effects. This quasi-static approximation is expected to be very accurate when the distance between the atomic system and the nanoparticle is subwavelength with respect to the atomic transition frequency $\omega_a$. Furthermore, our analysis assumes that the nanoparticle does not have a magnetic response $(\mu = \mu_0)$ so that it does not back-scatter appreciably the fields created by the atomic spin magnetic moment. In these conditions, the effective Hamiltonian is determined to leading order by [Eqs. (C8) and (C10)]:

$$h_{m,n} \approx -iA \frac{\alpha_0}{16\varepsilon_0 \pi^3 d^6} \text{tr}\left\{ \mathbf{Q}^{(m,n)} \cdot \left( \frac{\boldsymbol{\omega}_{\text{ef}}}{\omega_p} \times \mathbf{1} \right) \right\}, \qquad (9)$$

with $\alpha_0 = 4\pi R^3$, $\boldsymbol{\omega}_{\text{ef}} = \boldsymbol{\omega}_0 + 3(\boldsymbol{\omega}_0 \times \hat{\mathbf{z}}) \times \hat{\mathbf{z}}$, $\boldsymbol{\omega}_0 = \omega_0 \hat{\mathbf{u}} = \frac{-q}{m^*} \mathbf{B}_0$ is the oriented cyclotron frequency, and $A$ is the real-valued dimensionless parameter

$$A = \int_0^\infty d\xi \frac{1}{\left(3\xi(\xi+\Gamma) + \omega_p^2\right)^2} \frac{3\xi^2 \omega_p^3}{\omega_a^2 + \xi^2}. \qquad (10)$$

In the above $\mathbf{Q}^{(m,n)}$ should be understood as $\mathbf{Q}^{(m,n)} = \langle g_m | \hat{\mathbf{p}}_e^- \otimes \hat{\mathbf{p}}_e^+ | g_n \rangle$ (3×3 matrix) so that it is determined only by the electric dipole operator. The spin magnetic moment governs the cyclotron frequency vector $\boldsymbol{\omega}_0$ and introduces nonlinearity into the reduced master equation. It is important to note that neglecting the effects of the spin magnetic

moment would cause $h_{m,n}$ to vanish, except for a constant Lamb shift. The Lamb shift, responsible for the Casimir effect, has been excluded from the analysis (see Appendix C).

The described formalism applies to a generic time-reversal invariant two-level system, with the number of degenerate excited states arbitrary. For the case of a Kramers two-level system with excited states $|e_1\rangle$ and $|e_2\rangle$ linked by time-reversal symmetry (Fig. 1) $\mathbf{Q}^{(m,n)}$ are given by:

$$\mathbf{Q}^{(1,1)} = \boldsymbol{\gamma}_d \otimes \boldsymbol{\gamma}_d^* + \boldsymbol{\gamma}_c \otimes \boldsymbol{\gamma}_c^*, \qquad \mathbf{Q}^{(2,2)} = \boldsymbol{\gamma}_d^* \otimes \boldsymbol{\gamma}_d + \boldsymbol{\gamma}_c^* \otimes \boldsymbol{\gamma}_c, \tag{11a}$$

$$\mathbf{Q}^{(1,2)} = -\boldsymbol{\gamma}_d \otimes \boldsymbol{\gamma}_c + \boldsymbol{\gamma}_c \otimes \boldsymbol{\gamma}_d, \qquad \mathbf{Q}^{(2,1)} = \boldsymbol{\gamma}_d^* \otimes \boldsymbol{\gamma}_c^* - \boldsymbol{\gamma}_c^* \otimes \boldsymbol{\gamma}_d^*. \tag{11b}$$

Substituting these formulas into Eq. (9) one obtains the following explicit formulas for the Hamiltonian matrix elements:

$$h_{11} = -h_{22} = -A \frac{\alpha_0}{16\pi^3 \varepsilon_0 d^6} \frac{\boldsymbol{\omega}_{ef}}{\omega_p} \cdot \left( i\boldsymbol{\gamma}_d \times \boldsymbol{\gamma}_d^* + i\boldsymbol{\gamma}_c \times \boldsymbol{\gamma}_c^* \right), \tag{12a}$$

$$h_{12} = h_{21}^* = -iA \frac{\alpha_0}{8\pi^3 \varepsilon_0 d^6} \frac{\boldsymbol{\omega}_{ef}}{\omega_p} \cdot \left( \boldsymbol{\gamma}_c \times \boldsymbol{\gamma}_d \right). \tag{12b}$$

As seen, the interaction with the (nonreciprocal) environment leads to Lamb shifts with opposite signs for the two ground states, $h_{11} = -h_{22}$ (Zeeman type effect). Note that $h_{11}, h_{22}$ are always real-valued, whereas the anti-diagonal terms $h_{12}, h_{21}$ may be complex-valued. Furthermore, the above formulas confirm that $\mathbf{H}_{ef}$ is an Hermitian matrix, $\mathbf{H}_{ef} = \mathbf{H}_{ef}^\dagger$, as it should be to guarantee the unitary evolution of the atomic state. Importantly, the diagonal terms are nonzero only in presence of atomic chiral-type transitions.

It is underlined that $h_{mn}$ is proportional to the magnetic field $\mathbf{B}_0$ created by the spin magnetic moment. Clearly, for a fixed magnetic field $\mathbf{B}_0$ the Hamiltonian that describes the ground state is non-trivial. As a consequence, the ground state energy ($\mathcal{E} = \sum_{m,n} c_m^* h_{mn} c_n$) is state dependent. In particular, for a given atomic configuration the energy of the state $|g_1\rangle$ may be slightly different form the energy of $|g_2\rangle$ ($h_{11} \neq h_{22}$) due to the detuning arising from the interactions with the environment.

## C. Representation of the spin magnetic moment operator

From Eq. (4), the magnetic field $\mathbf{B}_0$ can be written as

$$\mathbf{B}_0 = \frac{\mu_0}{4\pi d^3} (3\hat{\mathbf{z}} \otimes \hat{\mathbf{z}} - \mathbf{1}) \cdot \begin{pmatrix} \mathbf{c}^* \cdot \mathbf{M}_1 \cdot \mathbf{c} \\ \mathbf{c}^* \cdot \mathbf{M}_2 \cdot \mathbf{c} \\ \mathbf{c}^* \cdot \mathbf{M}_3 \cdot \mathbf{c} \end{pmatrix}, \tag{13}$$

with $\mathbf{c} = (c_1 \ c_2)^T$ and $\mathbf{M}_i$ the matrix that represents the $i$-th component of the vector operator $\hat{\mathbf{m}}_s$ in the ground subspace basis ($|g_1\rangle, |g_2\rangle$). For simplicity, we shall suppose that $|g_1\rangle, |g_2\rangle$ can be identified with the spin up and spin down states as in Eq. (2). Then, $\mathbf{M}_i$ is proportional to the $i$-th Pauli matrix $\boldsymbol{\sigma}_i$:

$$\mathbf{M}_i = \frac{\hbar}{2} \frac{q}{m_e} \boldsymbol{\sigma}_i. \tag{14}$$

Here, $m_e$ is the free-electron mass, $q = -e$, and an electron $g$-factor of 2 is assumed. Within these assumptions the oriented cyclotron frequency $\boldsymbol{\omega}_0 = \frac{-q}{m^*} \mathbf{B}_0$ becomes:

$$\boldsymbol{\omega}_0 = \frac{\mu_0}{4\pi d^3} \frac{\hbar}{2} \frac{q^2}{m_e m^*} \begin{pmatrix} c_1^* c_2 + c_2^* c_1 \\ -i c_1^* c_2 + i c_2^* c_1 \\ 2(|c_2|^2 - |c_1|^2) \end{pmatrix}. \tag{15}$$

It is interesting to note that under a time reversal the atomic state $|\psi\rangle = c_1|g_1\rangle + c_2|g_2\rangle$ is transformed into $\mathcal{T}|\psi\rangle = -c_2^*|g_1\rangle + c_1^*|g_2\rangle$ [44]. Thus, a time reversal transforms the coefficients $(c_1, c_2)$ as $(c_1, c_2) \to (-c_2^*, c_1^*)$. Note that the time-reversal operator is antilinear. This property shows that the cyclotron frequency $\omega_0$ is flipped under a time-reversal ($\omega_0 \to -\omega_0$), as it should be.

Curiously, in absence of material loss both the atom and the environment are time-reversal invariant. Consistent with this property, the effective Hamiltonian is also time-reversal symmetric. This property still holds true even when the environment is lossy. To demonstrate this, we note that the time-reversal invariance requires that $\mathbf{H}_{ef}(\omega_0) = \sum_{m,n} h_{m,n}(\omega_0)|g_m\rangle\langle g_n|$ satisfies $\mathbf{H}_{ef}(\omega_0) = \mathcal{T}^{-1}\mathbf{H}_{ef}(-\omega_0)\mathcal{T}$ or equivalently $\mathbf{H}_{ef}(\omega_0) = -\mathcal{T}\mathbf{H}_{ef}(-\omega_0)\mathcal{T}$ [44]. As $(\mathcal{KT})^\dagger = -\mathcal{KT}$ with $\mathcal{K}$ the conjugation operator [44], this is still the same as $\mathbf{H}_{ef}(\omega_0) = \sum_{m,n} h_{mn}^*(-\omega_0)|\mathcal{T}g_m\rangle\langle \mathcal{T}g_n|$. Using now Eq. (12), $|g_2\rangle = \mathcal{T}|g_1\rangle$ and $|g_1\rangle = -\mathcal{T}|g_2\rangle$ one can easily check that the nonlinear Hamiltonian is indeed time-reversal invariant. In general, an effective Hamiltonian of the type $\mathbf{H}_{ef} = \mathbf{H}_{ef}(c_1, c_2)$ is time-reversal symmetric when:

$$\mathbf{H}_{ef}(c_1, c_2) = \begin{pmatrix} h_{11} & h_{12} \\ h_{21} & h_{22} \end{pmatrix}\bigg|_{(c_1,c_2)} = \begin{pmatrix} h_{22}^* & -h_{21}^* \\ -h_{12}^* & h_{11}^* \end{pmatrix}\bigg|_{(-c_2^*,c_1^*)}. \tag{16}$$

Examples of nonlinear time-reversal invariant classical systems are discussed in [53].

## V. Time-Crystal States and Spontaneous Symmetric Breaking

In the following, we demonstrate that the nonlinear quantum master equation can give rise to either time-crystal-like states or spontaneous symmetry breaking of time-reversal symmetry.

To this end, we consider a toy model for the Kramers two-level system that generalizes Eq. (3). Similar to the hydrogen-like model of Sect. II.A, one the atomic transitions is circularly polarized whereas the other transition is linearly polarized:

$$\boldsymbol{\gamma}_d = \frac{\gamma_d}{\sqrt{2}}(\hat{\mathbf{x}} \pm i\hat{\mathbf{y}}), \qquad \boldsymbol{\gamma}_c = \gamma_c \hat{\mathbf{z}}. \qquad (17)$$

However, here the complex amplitudes $\gamma_c$ and $\gamma_d$ are unconstrained. It can be shown that for the "−" sign the system is invariant under continuous rotations about the z-axis [44]. For the "+" sign there is a continuous rotation symmetry about the z-axis only when $\gamma_c = 0$. "Low-symmetry" scenarios may occur naturally in some molecular systems [45]. The sign ± determines the handedness of the dipolar transition $|e_1\rangle \rightarrow |g_1\rangle$. The handedness direction is formally defined as $\hat{\mathbf{s}}_{\text{dip},1} = \frac{i}{2|\boldsymbol{\gamma}_d|^2} \boldsymbol{\gamma}_d \times \boldsymbol{\gamma}_d^* = \pm\hat{\mathbf{z}}$. The handedness for the dipolar transition $|e_2\rangle \rightarrow |g_2\rangle$ is the opposite one $\hat{\mathbf{s}}_{\text{dip},2} = \mp\hat{\mathbf{z}}$.

It is worth noting that the transformation $(\gamma_d, \gamma_c) \rightarrow (\gamma_c, -\gamma_d)$ leaves the effective Hamiltonian unchanged (see Eq. (12)). Thus, the physics remains qualitatively the same when the direct transitions are linearly polarized and the crossed transitions are circularly polarized.

Substituting $\boldsymbol{\gamma}_c$ and $\boldsymbol{\gamma}_d$ in Eq. (12) and using $\boldsymbol{\omega}_{ef} = \boldsymbol{\omega}_0 + 3(\boldsymbol{\omega}_0 \times \hat{\mathbf{z}}) \times \hat{\mathbf{z}}$ and Eq. (15), one can readily show that the effective Hamiltonian elements are:

$$h_{11} = -h_{22} = -A\frac{\alpha_0}{16\pi^3\varepsilon_0 d^6}\frac{\omega_{0\perp}}{\omega_p}(\pm 2)\left(|c_2|^2 - |c_1|^2\right)|\gamma_d|^2, \tag{18a}$$

$$h_{12} = h_{21}^* = -iA\frac{\alpha_0}{8\pi^3\varepsilon_0 d^6}\frac{\omega_{0\perp}}{\omega_p}2\sqrt{2}i\gamma_d\gamma_c \begin{cases} c_1^*c_2, & +\text{sign} \\ -c_1c_2^*, & -\text{sign} \end{cases}. \tag{18b}$$

with $\omega_{0\perp} = \alpha_{e,fs}\frac{\hbar^2}{2m_e m^*}\frac{1}{c}\frac{1}{d^3}$. Hence, the effective Hamiltonian ($\mathbf{H}_{ef} = \mathbf{H}_{ef,\pm}$ with the $\pm$ sign determined by the handedness of the dipolar transition $|e_1\rangle \to |g_1\rangle$) reduces to:

$$\mathbf{H}_{ef,\pm}(c_1, c_2) = \pm \begin{pmatrix} \mathcal{E}_d\left(|c_1|^2 - |c_2|^2\right) & \mathcal{E}_{cr}c_a^*c_b \\ \mathcal{E}_{cr}^*c_a c_b^* & -\mathcal{E}_d\left(|c_1|^2 - |c_2|^2\right) \end{pmatrix}, \quad \text{with} \tag{19a}$$

$$\mathcal{E}_d = A\frac{R^3}{2\pi^2\varepsilon_0 d^6}\frac{\omega_{0\perp}}{\omega_p}|\gamma_d|^2 \quad \text{and} \quad \mathcal{E}_{cr} = 2\sqrt{2}\frac{\gamma_c}{\gamma_d^*}\mathcal{E}_d. \tag{19b}$$

Note that $\mathcal{E}_d, \mathcal{E}_{cr}$ are constants independent of the atomic state. In the above, $(a,b) = (1,2)$ for the "+" sign case, whereas $(a,b) = (2,1)$ for the "−" sign case.

For a given initial state $\mathbf{c} = (c_1, c_2)$, the system dynamics may be characterized by solving the nonlinear Schrödinger equation (7). As previously mentioned, since $\mathbf{H}_{ef} = \mathbf{H}_{ef}^\dagger$ the time evolution is unitary such that $\mathbf{c}^* \cdot \mathbf{c} = 1$ at all time instants. Each atomic state is associated with a well-defined spin vector direction given by $\hat{\mathbf{S}} = (\mathbf{c}^* \cdot \sigma_x \cdot \mathbf{c}, \mathbf{c}^* \cdot \sigma_y \cdot \mathbf{c}, \mathbf{c}^* \cdot \sigma_z \cdot \mathbf{c})$. Thus, it is intuitive to represent the system state on the Bloch sphere using the mapping $(c_1, c_2) \to \hat{\mathbf{S}}$. The North and South poles of the Bloch sphere correspond to the states $|g_1\rangle$ and $|g_2\rangle$, respectively. As the spin vector flips under a time reversal ($(c_1, c_2) \rightarrow (-c_2^*, c_1^*)$), the time reversed system state is represented by the antipodal point of the Bloch sphere.

We recall that $|g_1\rangle$ and $|g_2\rangle$ are the states with electron spin oriented along the +z-direction (−z-direction). Thus, the Hamiltonian $\mathbf{H}_{ef,+}$ models a system such that the handedness of the chiral transition $|e_1\rangle \rightarrow |g_1\rangle$ ($\hat{\mathbf{s}}_{dip,1} = +\hat{\mathbf{z}}$) matches the handedness of the electron spin (spin up). Likewise the handedness of the transition $|e_2\rangle \rightarrow |g_2\rangle$ ($\hat{\mathbf{s}}_{dip,2} = -\hat{\mathbf{z}}$) also matches the handedness of the corresponding electron spin (spin down). In contrast, the Hamiltonian $\mathbf{H}_{ef,-}$ models a system such that the handedness of the chiral dipolar transition is opposite to the handedness of the electron spin.

## A. Chiral polarized transitions dominate

In the first set of examples, we suppose that the chiral dipolar transitions dominate $|\gamma_d| \gg |\gamma_c|$ as in the Jahn-Teller $X_3$ molecule [45]. To begin with, let us ignore the linear polarized transitions and assume that $\gamma_c = 0$. In that case, $h_{12} = h_{21} = 0$ and the solution of the nonlinear Schrödinger equation [Eq. (7)] can be found analytically as follows:

$$c_1(t) = c_{1,t=0} e^{-i \frac{\pm \mathcal{E}_d}{\hbar} \left( |c_1|^2 - |c_2|^2 \right) t}, \qquad c_2(t) = c_{2,t=0} e^{+i \frac{\pm \mathcal{E}_d}{\hbar} \left( |c_1|^2 - |c_2|^2 \right) t}. \qquad (20)$$

Note that $|c_1|$ and $|c_2|$ are time independent and determined by the initial conditions ($c_{n,t=0} = |c_n| e^{i\phi_{0n}}$, n=1,2). The energy expectation is determined by quadratic form $\mathbf{c}^* \cdot \mathbf{H}_{ef,\pm}(\mathbf{c}) \cdot \mathbf{c} = \pm \mathcal{E}_d \left( |c_1|^2 - |c_2|^2 \right)^2$. In particular, for $\mathbf{H}_{ef-}$ the spin up and the spin down states are global minima of the quadratic form $\mathbf{c}^* \cdot \mathbf{H}_{ef}(\mathbf{c}) \cdot \mathbf{c}$, whereas for $\mathbf{H}_{ef+}$ they are

global maxima. When $\gamma_c = 0$ the energy is conserved along the orbit. In the general case, $\mathbf{c}^* \cdot \mathbf{H}_{ef}(\mathbf{c}) \cdot \mathbf{c}$ does not need to be constant of motion because some energy can be exchanged between the atom and the environment (see Sect. V.D).

The spin vector expectation associated with the orbit is

$$\hat{\mathbf{S}} = \left( 2|c_1||c_2|\cos(\omega_m t + \delta\phi), 2|c_1||c_2|\sin(\omega_m t + \delta\phi), |c_1|^2 - |c_2|^2 \right), \tag{21}$$

where $\delta\phi = \phi_{02} - \phi_{01}$ and

$$\omega_m = \frac{h_{11} - h_{22}}{\hbar} = \frac{\pm 2\mathcal{E}_d}{\hbar}\left(|c_1|^2 - |c_2|^2\right). \tag{22}$$

Clearly, notwithstanding the continuous time translation symmetry of both the microscopic and effective Hamiltonians, the atomic state varies periodically in time. Excluding an overall phase factor of the wave function with no physical consequences, the time period of the wave function is $T_{spin} = \frac{2\pi}{|\omega_m|}$, which is the same as the time period of the spin vector. Thus, the "ground" is a time-crystal-like state.

The spontaneous symmetry breaking of the time translation symmetry is rooted in the precession of the spin vector around the $z$-axis. Figure 2 represents the dynamics of a generic pure state in the Bloch sphere. The density matrix ($\hat{\rho}_S(t) = \sum_{i,j} \rho_{ij} |g_i\rangle\langle g_j|$) depends on the coordinates $c_1, c_2$ as $\rho_{ij} = c_i c_j^*$. We choose an initial state such that $\rho_{11,t=0} = 0.7$ (blue dot in Figs. 2a and 2b). For reference, the purple dots in the figures represent the states $|g_1\rangle$ (north pole) and $|g_2\rangle$ (south pole). It can be seen in Figs. 2a and 2b, the direction of the precession motion depends on the considered Hamiltonian $\mathbf{H}_{ef,\pm}$.

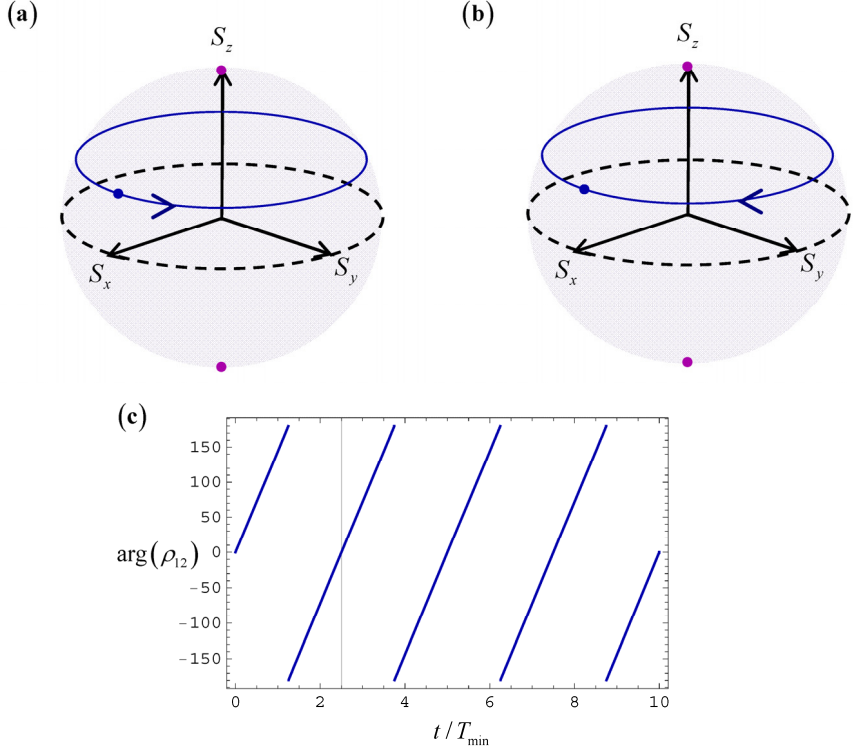

**Fig. 2** Representation of the trajectory of the spin vector (blue curve) in the Bloch sphere for the Hamiltonian **a)** $\mathbf{H}_{ef,+}$ and **b)** $\mathbf{H}_{ef,-}$. The blue arrow indicates the direction of growing time. The initial atomic state (blue dot) is a pure state with $\rho_{11} = 0.7$. The dashed black circle represents the equator line of the Bloch sphere and the purple dots the North and South poles (spin up and spin down states). **c)** Phase of $\rho_{12}$ as a function of the normalized time for the Hamiltonian $\mathbf{H}_{ef,-}$. The dashed vertical line marks the theoretical precession period.

For the Hamiltonian $\mathbf{H}_{ef,+}$ (Fig. 2a) the precession angular velocity ($\boldsymbol{\omega}_m$) is locked to the direction of the spin vector in the same manner as for a classical spinning top: $\boldsymbol{\omega}_m \sim (\hat{\mathbf{S}} \cdot \hat{\mathbf{z}})\hat{\mathbf{z}}$. On the other hand, for the Hamiltonian $\mathbf{H}_{ef,-}$ (Fig. 2b) the direction of the precession angular velocity is opposite to the direction of the spin vector $\boldsymbol{\omega}_m \sim -(\hat{\mathbf{S}} \cdot \hat{\mathbf{z}})\hat{\mathbf{z}}$. Thus, the precession motion relative to the direction of the spin vector depends if the handedness of the chiral transitions matches the handedness of the atomic spin or not.

Note that when the pure states lies in the equator of the Bloch sphere ($\rho_{11,t=0} = \rho_{22,t=0} = 1/2$) the time dynamics is trivial and there is no spin precession.

The shortest time-crystal period ($T_{\min}$) occurs when the state is near the poles of the Bloch sphere and is given by $T_{\min} = \pi\hbar/\mathcal{E}_d$. To give an idea of the relevant time scale of the problem, consider a semiconductor nanoparticle with $\omega_p/2\pi = 1\,\text{THz}$, $\Gamma = 0.1\omega_p$, $m^* = 0.001 m_e$ and radius $R = d/2$. The nanoparticle is separated from the atomic system by $d = 5nm$. Supposing that the transition electric dipole moment is on the order of $\gamma_d = 100$ Debye and that the transition frequency is $\omega_a = 50\omega_p$ one finds that $1/T_{\min} = 63\,\text{kHz}$. So the spin precession is a quite slow process which justifies the Born approximation and the assumption that the environment changes adiabatically with atomic spin magnetic moment. Note that $1/T_{\min}$ scales with $\gamma_d^2$ and so for smaller values of the electric dipole the precession frequency is even smaller. Figure 2c represents the phase of $\rho_{12} = c_1 c_2^*$ for the Hamiltonian $\mathbf{H}_{\text{ef},-}$. After a full time crystal cycle the phase changes by $2\pi$. The time crystal period (dashed vertical line) is $T_{\text{spin}} = T_{\min}/\left||c_1|^2 - |c_2|^2\right|$ and depends on the initial state.

### *B. Spontaneous symmetry breaking of time reversal symmetry*

Next, we discuss the impact of weak linearly polarized crossed transitions on the time crystal dynamics $|\gamma_d| \gg |\gamma_c| \neq 0$ (Fig. 3). For $|\gamma_c| \neq 0$ and $|\gamma_d| \neq 0$ the anti-diagonal elements of the effective Hamiltonian are nonzero. Hence, there is a nontrivial coupling between the two electron spins and the nonlinear equations of motion cannot be

integrated analytically. For the Hamiltonian $\mathbf{H}_{ef,+}$ the effect of the crossed transitions is mild (Fig. 3a). The perturbation creates some wobbling with respect to a purely circular orbit, but always leads to closed orbits in the Bloch sphere. The situation for the Hamiltonian $\mathbf{H}_{ef,-}$ is far more interesting (Figs. 3bi-iii). In this case, a small $\gamma_c$ can completely disrupt the time crystal cycle when the phase of the parameter $\gamma_c$ is different from the phase of $\gamma_d^*$. Specifically, independent of the initial (pure) state the orbit of the spin vector evolves towards the South (North) pole of the sphere when $0 < \arg(\gamma_c / \gamma_d^*) < \pi$ ($-\pi < \arg(\gamma_c / \gamma_d^*) < 0$), respectively (Figs. 3bii and 3biii). For $\arg(\gamma_c / \gamma_d^*) = 0$ or $\arg(\gamma_c / \gamma_d^*) = \pi$, the orbit remains circular and closed (Fig. 3bi).

Thereby, when the handedness of the atomic spin does not match the handedness of the chiral dipole transition the time-crystal dynamics is typically suppressed after a transient process that occurs on a time scale on the order of $T_{min} |\gamma_d| / |\gamma_c|$. Remarkably, independent of the initial preparation of the system, for large $t$ the ground-state is either the spin down state ($0 < \arg(\gamma_c / \gamma_d^*) < \pi$) or the spin up state ($-\pi < \arg(\gamma_c / \gamma_d^*) < 0$). In such circumstances the stable ground state is not invariant under a time-reversal. Thus, the ground state has less symmetry than the system Hamiltonian, in other words, the time-reversal symmetry is spontaneously broken. It is underlined that $\mathbf{H}_{ef,-}$ is time-reversal symmetric.

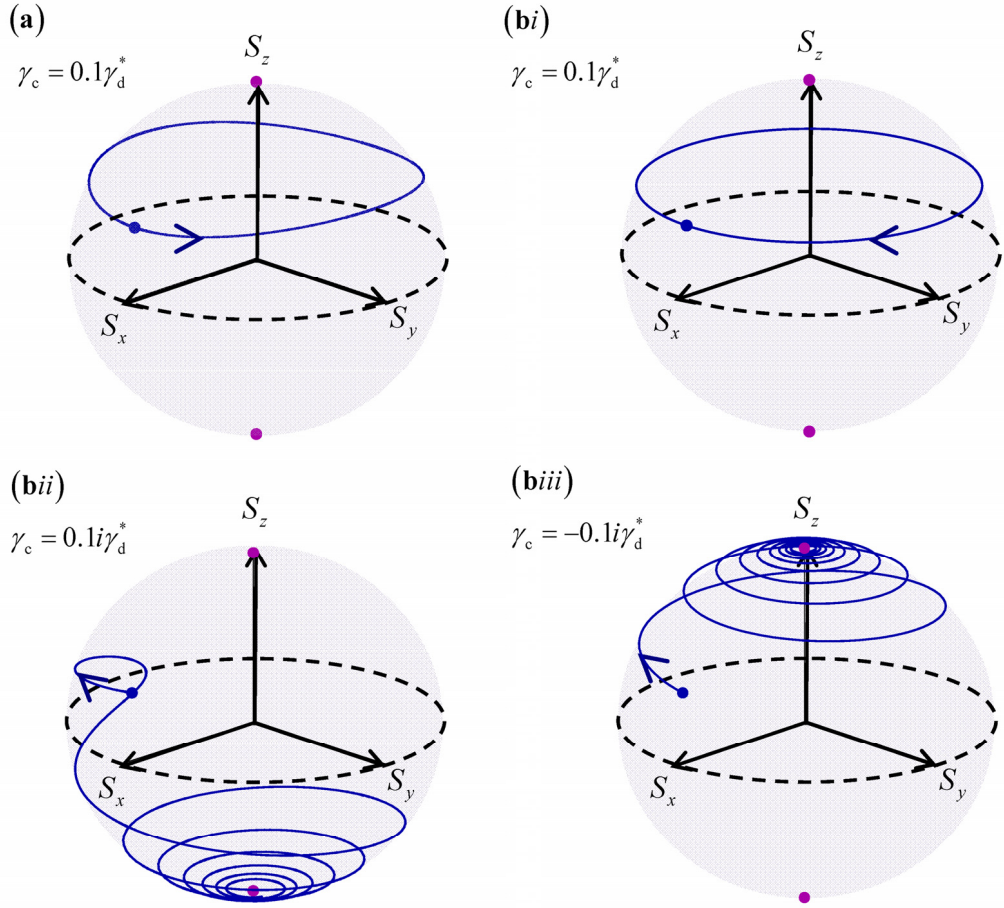

**Fig. 3** Representation of the trajectory of the spin vector in the Bloch sphere for the Hamiltonian **a)** $\mathbf{H}_{ef,+}$ and **b)** $\mathbf{H}_{ef,-}$. Here, it is supposed that $|\gamma_c| = 0.1|\gamma_d|$ and the rest of the parameters are as in Fig. 2.

The different dynamics of the Hamiltonians $\mathbf{H}_{ef,\pm}$ can be formally justified using the Lindblad master equation that controls the time dynamics of $\rho_{11} = |c_1|^2$. From Appendix E [Eq. (E1)], one finds that $\partial_t \rho_{11} = \frac{2}{\hbar} \text{Im}\{h_{12}\rho_{21}\}$. Thus, taking into account that for a pure state $\rho_{21} = c_2 c_1^*$ and using Eq. (19) it follows that:

$$\partial_t \rho_{11} = \sqrt{2} \frac{4\mathcal{E}_d}{\hbar} \begin{cases} +\mathrm{Im}\left\{\dfrac{\gamma_c}{\gamma_d^*}\left(c_1^* c_2\right)^2\right\}, & \text{for } \mathbf{H}_{\text{ef},+} \\ -|c_1 c_2|^2 \mathrm{Im}\left\{\dfrac{\gamma_c}{\gamma_d^*}\right\}, & \text{for } \mathbf{H}_{\text{ef},-} \end{cases} \tag{23}$$

As seen, for the Hamiltonian $\mathbf{H}_{\text{ef},-}$ the sign of $\partial_t \rho_{11}$ is independent of time and is strictly locked to the sign of $-\mathrm{Im}\{\gamma_c/\gamma_d^*\}$. In agreement with Fig. 3, $\rho_{11}$ is a strictly increasing (decreasing) function of time when $\mathrm{Im}\{\gamma_c/\gamma_d^*\}<0$ ($\mathrm{Im}\{\gamma_c/\gamma_d^*\}>0$). In contrast, for the Hamiltonian $\mathbf{H}_{\text{ef},+}$ the sign of $\partial_t \rho_{11}$ depends on the system state and oscillates between positive and negative values along the time-crystal orbit, explaining the wobbling motion observed in Fig. 3bi.

To understand the compatibility of the spontaneously broken symmetry with the $\mathcal{T}$-invariant effective Hamiltonian, we depict in Fig. 4a the full trajectory of the spin vector expectation in the range $-\infty < t < +\infty$ for the same example as in Fig. 3biii ($\gamma_c = -0.1 i \gamma_d^*$). Independent of the initial state at $t=0$, the orbit for $t \to -\infty$ starts always at the South pole of the Bloch sphere ("repeller point") and ends always at the North pole of the sphere ("attractor point"). Attractor and repeller points are well-known in the theory of nonlinear dynamics [54]. The time reversal symmetry of the Hamiltonian implies that if $\hat{\mathbf{S}}(t)$ is the trajectory associated with a certain initial state, then $-\hat{\mathbf{S}}(-t)$ is the trajectory associated with the corresponding time-reversed initial state. Then, the existence of an orbit attractor for $t \to +\infty$ forcibly implies that the existence of an orbit repeller for $t \to -\infty$. Furthermore, the attractor and the repeller must be linked by the time-reversal operator, in agreement with the numerical results in Fig. 4. Note that attractor and the repeller points are forcibly eigenstates of the nonlinear Hamiltonian:

$\mathbf{H}_{\text{ef}}(\mathbf{c}) \cdot \mathbf{c} = \mathcal{E}\mathbf{c}$.

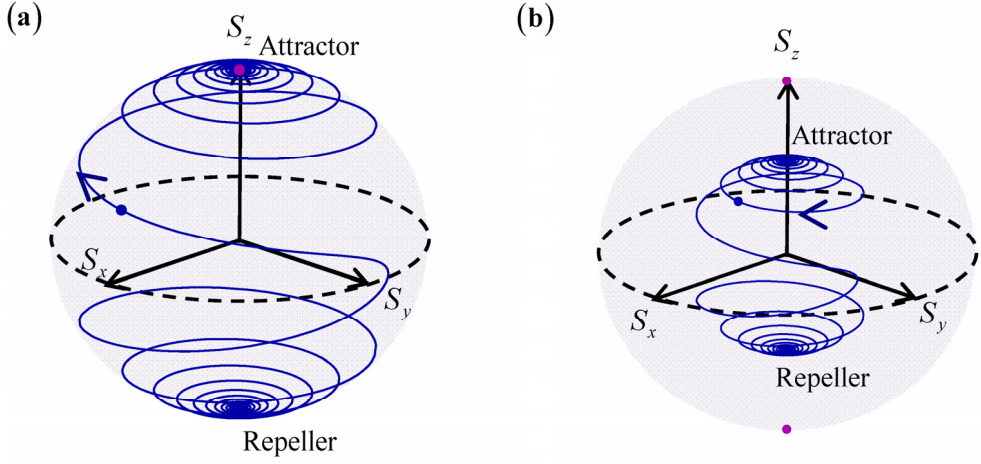

**Fig. 4** Representation of the trajectory of the spin vector in the Bloch sphere for the Hamiltonian $\mathbf{H}_{\text{ef},-}$ with $\gamma_c = -i0.1\gamma_d^*$ for a **a)** pure state and for **b)** a mixed state. For an initial pure state all the trajectories converge to the attractor (repeller) when $t \to \infty$ ($t \to -\infty$).

The spontaneously broken symmetry can, in principle, be exploited to generate nonreciprocal light-matter interactions without an external magnetic field bias. For example, in Fig. 4a, the vacuum fluctuations anchor the atomic spin vector along the +z-axis, creating a nonreciprocal response in both the atom and nanoparticle electric polarizabilities. Since the spin-up state is the only stable orientation of the spin vector, the nonreciprocal response is expected to remain resilient to sufficiently weak external perturbations.

It is also relevant to discuss the time dynamics of mixed states. The time evolution of the mixed states is determined by numerically solving the reduced master equation $\partial_t \hat{\rho}_S = -i\frac{1}{\hbar}\left[\hat{H}_{\text{ef}}, \hat{\rho}_S\right]$ (see Eq. (E1) and the associated discussion). Now, the spin vector expectation is written in terms of the density matrix elements as $\hat{\mathbf{S}} = (\rho_{12} + \rho_{21}, i\rho_{12} - i\rho_{21}, \rho_{11} - \rho_{22})$. For mixed states the amplitude of the spin vector may be less than the unity ($|\hat{\mathbf{S}}| \leq 1$). Thus, in general $\hat{\mathbf{S}}$ is a point interior to the Bloch sphere. Figure 4b shows the time evolution of an initial mixed state with $\rho_{11,t=0} = 0.7$, $\rho_{22,t=0} = 0.3$ and $\rho_{12,t=0} = 0.18$. As seen, also in this case the attractor (repeller) states have a spin vector expectation oriented towards the +z-direction (−z-direction), but they are not pure states. For this example, any mixed state with $\rho_{12} = 0$ and $\rho_{11} > \rho_{22}$ is a stable attractor point. Thus, there is a continuous family of attractors that are mixed states and a single pure attractor state.

### C. Linear polarized transitions dominate

Next, we consider the scenario where the linearly polarized dipole transitions dominate $0 \neq |\gamma_d| \ll |\gamma_c|$. Figure 5 shows the time evolution of pure states for an atomic system with $\gamma_c = -10i\gamma_d^*$. Comparing with Fig. 3, one sees that the Hamiltonian $\mathbf{H}_{ef,+}$ still supports time-crystal states (Fig. 5a) but the orbits now cross both the North and South hemispheres of the Bloch sphere, rather than being confined to one of the hemispheres (Fig. 3a). Thus, now the inversion $\rho_{11} - \rho_{22}$ may vary significantly in each time-crystal cycle. On the other hand, for the Hamiltonian $\mathbf{H}_{ef,-}$ one sees that similar to Fig. 3biii the orbit is dragged to the attractor point on the North pole of the Bloch sphere. However, the attraction is so strong that the precession around the North pole is not detectable anymore. We recall that the time scale associated with the spin precession is on the order of $T_{\min}$ whereas the time scale associated with the relaxation to the attractor

point is on the order of $T_{\min}|\gamma_d|/|\gamma_c|$ [Eq. (23)]. Consequently, the latter process dominates when $|\gamma_c| \gg |\gamma_d|$ as in the present example.

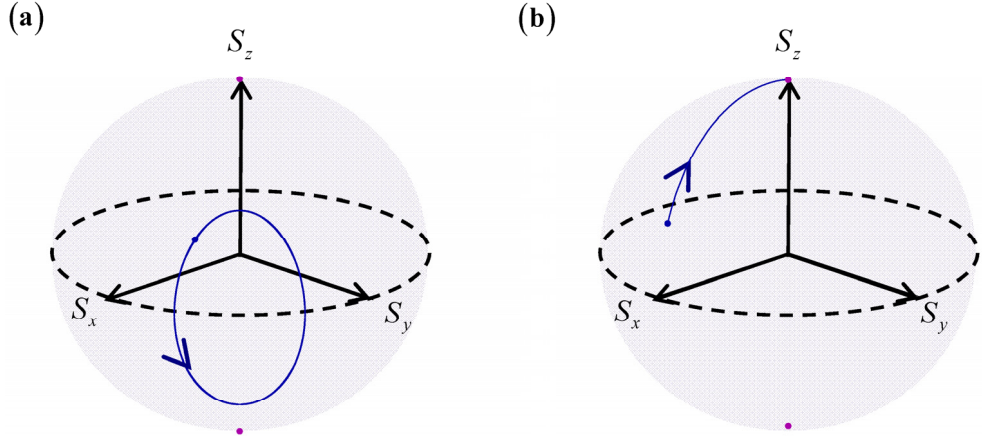

**Fig. 5** Representation of the trajectory of the spin vector in the Bloch sphere for the Hamiltonian **a)** $\mathbf{H}_{\text{ef},+}$ and **b)** $\mathbf{H}_{\text{ef},-}$ and a pure initial state. Here, $\gamma_c = -i10\gamma_d^*$ so that linearly polarized transitions dominate. The initial atomic state (blue dot) is a pure state with $\rho_{11} = 0.7$.

### D. Closed orbits and conservative dynamics

It is rather remarkable that the orbits associated with the Hamiltonian $\mathbf{H}_{\text{ef},+}$ are exactly closed and periodic. Next, we demonstrate that this property arises from a hidden conservative dynamics. To this end, first we note that the energy expectation of the Hamiltonian $\mathbf{H}_{\text{ef},+}$ can be explicitly written as follows:

$$\langle \mathbf{H}_{\text{ef},+} \rangle \equiv \mathbf{c}^* \cdot \mathbf{H}_{\text{ef},+}(\mathbf{c}) \cdot \mathbf{c} = \mathcal{E}_d \left( c_1 c_1^* - c_2 c_2^* \right)^2 + \mathcal{E}_{\text{cr}} \left( c_1^* \right)^2 \left( c_2 \right)^2 + \mathcal{E}_{\text{cr}}^* \left( c_2^* \right)^2 \left( c_1 \right)^2. \qquad (24)$$

Thus, $\langle \mathbf{H}_{\text{ef},+} \rangle$ may be regarded as an analytical function of the variables $c_1, c_1^*, c_2, c_2^*$. Interestingly, it can be readily verified that for the $\mathbf{H}_{\text{ef},+}$ Hamiltonian the nonlinear-Schrödinger equation (7) is formally equivalent to the following set of equations:

$$i2\hbar \frac{d}{dt} c_i = \frac{\partial}{\partial c_i^*} \langle \mathbf{H}_{ef,+} \rangle, \qquad i2\hbar \frac{d}{dt} c_i^* = -\frac{\partial}{\partial c_i} \langle \mathbf{H}_{ef,+} \rangle, \quad (i=1,2). \tag{25a}$$

This equivalence implies that the nonlinear Schrödinger equation generates precisely the same dynamics as a classical Hamiltonian function $\langle \mathbf{H}_{ef,+} \rangle$, with $c_i, c_i^*$ canonical conjugate variables! In particular, the energy expectation $\langle \mathbf{H}_{ef,+} \rangle$ is a constant of motion because $\frac{d}{dt} \langle \mathbf{H}_{ef,+} \rangle = \sum_i \frac{\partial \langle \mathbf{H}_{ef,+} \rangle}{\partial c_i} \frac{dc_i}{dt} + \frac{\partial \langle \mathbf{H}_{ef,+} \rangle}{\partial c_i^*} \frac{dc_i^*}{dt} = 0$, resulting in orbits that are exactly closed and periodic. In fact, these orbits are determined by the curves of the Bloch sphere generated by the condition $\langle \mathbf{H}_{ef,+} \rangle = const.$.

In contrast, the dynamics of $\mathbf{H}_{ef,-}$ is conservative only when the parameter $\mathcal{E}_{cr}$ is real-valued. For a complex-valued $\mathcal{E}_{cr}$, the energy expectation $\langle \mathbf{H}_{ef,-} \rangle$ is not conserved, and the time-dynamics lead to the system state evolving towards minimizing $\langle \mathbf{H}_{ef,-} \rangle$. These properties underscore the fundamental differences between the two Hamiltonians.

### E. *Radiation effect*

As seen in the previous sub-sections, the interaction of the spin magnetic moment with the nanoparticle may lead to the formation of periodic time-crystal-like states with a characteristic period on the order of $\hbar \pi / \mathcal{E}_d$. The time-crystal cycle is rooted in the precession of the spin magnetic moment. Hence it is natural to wonder if radiation effects can lead to a relaxation of the oscillation.

In order to investigate this, in Appendix E we incorporate in our analysis the effect of the radiation loss due to the atomic spin magnetic moment precession. The analysis is done in a perturbative manner and for simplicity assumes that the chiral dipolar transitions dominate ($|\gamma_d| \gg |\gamma_c|$). It leads to a modified Lindblad master equation of the form $\partial_t \hat{\rho}_S = -i\frac{1}{\hbar}\left[\hat{H}_{ef}, \hat{\rho}_S\right] + \mathcal{L}_m \hat{\rho}_S$. The last term ($\mathcal{L}_m \hat{\rho}_S$) describes a parametric relaxation process due to the emission of radiation, analogous to the standard spontaneous emission decay from an excited state to the ground state. The associated decay rate is given by $\Gamma_{sp} = g_{sp} \frac{2\mathcal{E}_d}{\hbar} |\rho_{11} - \rho_{22}|^3$ [Eq. (E7)] with $g_{sp} = \alpha_{e,fs} \frac{8}{3}\left(\frac{\mathcal{E}_d}{m_e c^2}\right)^2$ a dimensionless parameter. The parametric relaxation process is controlled by the spin precession frequency $\omega_m = \frac{2h_{11}}{\hbar} = \frac{\pm 2\mathcal{E}_d}{\hbar}(\rho_{11} - \rho_{22})$ [Eq. (22)]. When $\omega_m > 0$ the radiation effects tends to decrease the $\rho_{11}$ population, whereas when $\omega_m < 0$ the system evolves to increase the $\rho_{11}$ population. Note that the sign of $\omega_m$ depends on the system state and on the effective Hamiltonian ($\mathbf{H}_{ef,\pm}$).

To begin with, we estimate the peak value of the decay rate using $\Gamma_{sp} \sim g_{sp} \frac{2\mathcal{E}_d}{\hbar}$. Considering the same structural parameters as in section V.A, we get $g_{sp} = 1.3 \times 10^{-33}$, which corresponds to a decay rate on the order of $\Gamma_{sp} \sim 1.3 \times 10^{-33} \times 2\pi \times 63\,\text{kHz}$. The associated decay time is $1/\Gamma_{sp} \sim 6 \times 10^{19}$ years, many orders of magnitude larger than the age of the universe! So the effect of the magnetic dipole radiation is totally negligible, and for all practical purposes, within the validity of the model, the lifetime of the time crystal states is infinite. It is relevant to point out that in the quasi-static approximation considered here the radiative decay due to the magnetic dipole precession is not influenced by the dissipation in the plasmonic nanoparticle. Indeed, in the quasi-static limit the electric and magnetic fields are effectively decoupled and so the magnetic dipole is not influenced by the loss in the permittivity function.

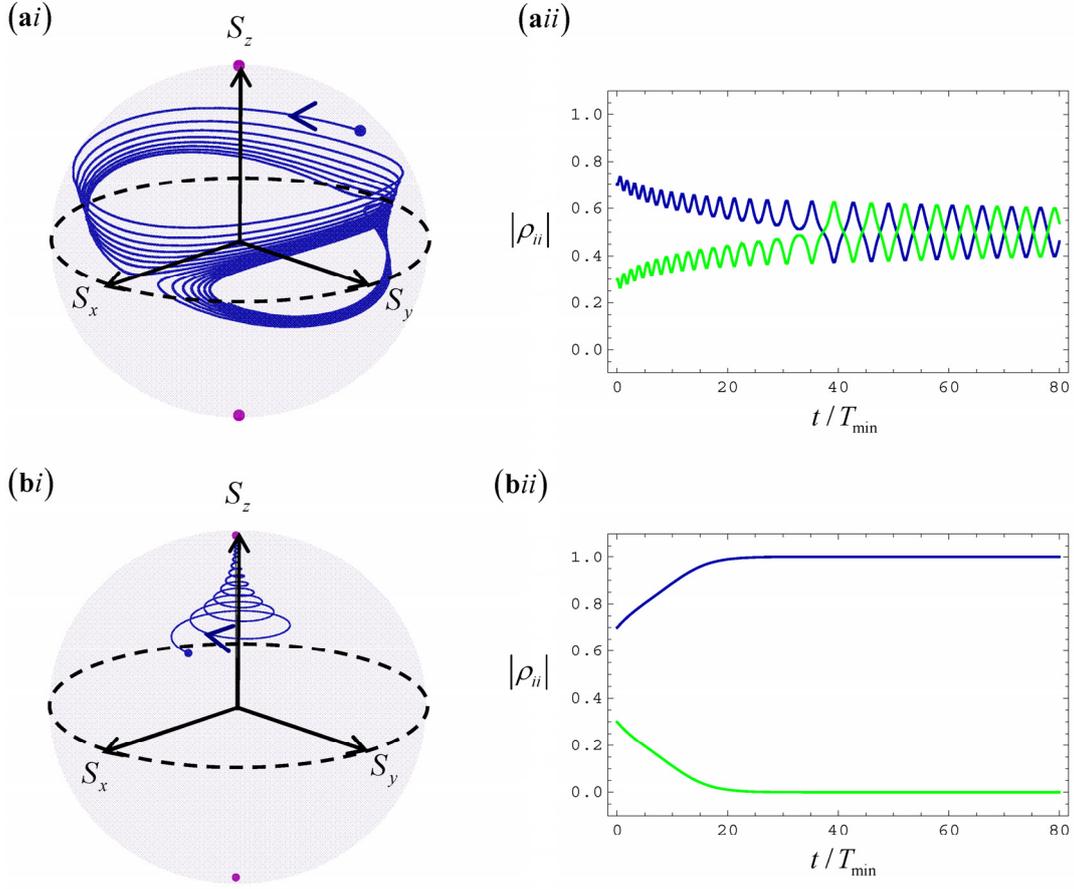

**Fig. 6** *i*) Effect of radiation loss ($g_{sp} = 0.05$) on the trajectory of the spin vector expectation in the Bloch sphere for **a)** $\mathbf{H}_{ef,+}$, $\gamma_c = 0.1\gamma_d^*$, with an initial pure state. **b)** $\mathbf{H}_{ef,-}$, $\gamma_c = -i0.1\gamma_d^*$ and an initial mixed state. *ii*) Plot of $\rho_{11}$ (blue line) and $\rho_{22}$ (green line) as a function of time.

Yet, it is interesting to assess how the dissipation affects the system dynamics. To this end, we consider an unrealistically large decay rate corresponding to $g_{sp} = 0.05$. Figure 6a shows the effect of radiation loss on the time evolution of the spin vector expectation for a time crystal state of the Hamiltonian $\mathbf{H}_{ef,+}$. As seen, after a few cycles the orbit approaches the plane $S_z = 0$, where $\rho_{11} = \rho_{22} = 0.5$ (see Fig. 6a*ii*). It can be seen that the orbit is slightly interior to the Bloch sphere, indicating the formation of a mixed state. For comparison, the time crystal orbit without including the decay rate $\Gamma_{sp}$ is shown in Fig. 3a. The observed behavior can be explained as follows. For the Hamiltonian $\mathbf{H}_{ef,+}$ the spin frequency $\omega_m$ is positive in the North hemisphere of the Bloch sphere. Thus, the radiation loss acts to decrease the $\rho_{11}$ population, and thereby makes the system state approach the equatorial line. If the system state crosses the equatorial line to the South hemisphere, then $\omega_m$ becomes negative and the radiation loss acts to increase the $\rho_{11}$ population. Thus, for $t \to \infty$ the system state is some point in the $S_z = 0$ plane. The convergence is rather slow because the decay rate ($\Gamma_{sp} \sim |\rho_{11} - \rho_{22}|^3$) approaches zero when $\rho_{11} = \rho_{22} = 0.5$.

The effect of radiation loss for the Hamiltonian $\mathbf{H}_{ef,-}$ is qualitatively different. To illustrate this, we depict in Fig. 6b the orbit of the spin vector expectation for the same initial mixed state as in Fig. 4a, but considering the effects of radiation loss modeled by $g_{sp} = 0.05$ (note that here we only show the orbit calculated for $t > 0$). As seen, different from Fig. 4a, the orbit converges asymptotically to the North pole of the Bloch sphere. The justification is that for the Hamiltonian $\mathbf{H}_{ef,-}$ the spin frequency $\omega_m$ is negative when the system state lies in the North hemisphere and hence the radiation loss acts to increase the $\rho_{11}$ population, i.e., drags the system state to the North pole. In general, for $\mathbf{H}_{ef,-}$ the effect of radiation loss is to drag the system state to the closest pole, independent if the state is pure or mixed. However, it is important to keep in mind that the radiation loss relaxation competes with the much stronger and dominant relaxation due to the "attractor" point. Thus, the overall effect of the radiation loss is to guarantee that there is a single stable attractor point (spin up state in the present example). In other words, there are no stable mixed states in the presence of radiation loss.

It is relevant to point out that both for $\mathbf{H}_{ef,+}$ and $\mathbf{H}_{ef,-}$ the asymptotic ground state in the presence of radiation loss minimizes the atomic energy expectation, $\mathbf{c}^* \cdot \mathbf{H}_{ef,\pm}(\mathbf{c}) \cdot \mathbf{c} \approx \pm \mathcal{E}_d \left( |c_1|^2 - |c_2|^2 \right)^2$ for $|\gamma_d| \gg |\gamma_c|$.

## VI. Summary

In summary, we theoretically studied how the degeneracy of the atomic ground state and the intrinsic spin magnetic moment of the electron can result in either a spontaneous symmetry breaking of time-translation symmetry or a spontaneous symmetry breaking of the time-reversal. We argued that the magnetic field created by the electron intrinsic magnetic moment can create a strong coupling between the environment and the atom. Using a parametric Born-Markov approximation, we have derived a reduced quantum master equation that describes the time evolution of the ground-state as a function of time. We demonstrated that the ground-state dynamics can be described by a nonlinear Schrödinger equation. The corresponding effective Hamiltonian is Hermitian and exhibits time-reversal symmetry.

Our analysis focused on a particular toy model that describes a system with circularly polarized (chiral) and linearly polarized atomic transitions. We found out that when the dipolar transitions are purely chiral, the time translation symmetry is spontaneously broken, and the system can support time-crystal states. Both hydrogen-like systems with a strong spin-orbit coupling or the Jahn-Teller molecular systems $X_3$ can be interesting platforms to investigate these time-crystal states [45]. In these states, the spin vector undergoes complete precession around the symmetry axis in each time cycle. The direction of the precession angular velocity depends on whether the handedness of a chiral dipolar transition matches the handedness of the spin vector (system $\mathbf{H}_{ef,+}$), or not (system $\mathbf{H}_{ef,-}$).

We have shown that in the latter case ($\mathbf{H}_{ef,-}$), the time-crystal state can be strongly influenced by linearly polarized transitions. These transitions cause the relaxation of the time-crystal cycle towards a well-defined attractor point in the Bloch-sphere. As a result, time-reversal symmetry of the system is spontaneously broken, leading to the formation of a well-defined ground state which lacks time-reversal symmetry. The system exhibits nonreciprocal electromagnetic response in this stable ground state, which could be beneficial for nanophotonic applications such as the design of electromagnetic isolators on the nanoscale without external bias. The nonreciprocal response originates from the self-bias provided by the vacuum fluctuations, which is responsible for the spontaneous symmetry breaking.

In contrast, the time crystal states supported by the Hamiltonian $\mathbf{H}_{ef,+}$ are resilient to the influence of linearly polarized dipolar transitions. Additionally, we conducted a perturbative study of the relaxation of the time-crystal state caused by the radiation emitted during each precession cycle. Our analysis shows that this effect is negligible, and that for all practical purposes the time crystal state lifetime is infinite.

Our theory demonstrates that the ground state physics of chiral-atomic systems can be remarkably intricate and offer intriguing applications in nonreciprocal nanophotonics. We hope that our theoretical findings will inspire further studies based on realistic physical platforms.

**Acknowledgements:** This work is supported in part by the Institution of Engineering and Technology (IET), by the Simons Foundation, and by Fundação para a Ciência e a Tecnologia and Instituto de Telecomunicações under project UIDB/50008/2020.

# Appendix A: The quantum master equation

Here, we derive the quantum master equation for an atom interacting with a structured environment, e.g., gyro-electric nanoparticle.

*A. Quantized electromagnetic field*

As discussed in Sect. II.B, in our problem the response of the environment is nonlinear at a microscopic level. To circumvent this difficulty (i.e., the quantization of a nonlinear bosonic field), we follow the standard approach and model the quantized field using the macroscopic (linearized) permittivity response given by Eq. (5). Due to this reason in our analysis the field operators will depend parametrically on the magnetic field created by the spin magnetic moment. This is property is implicit in all the discussions of Appendix A.

The environment degrees of freedom are modeled by a continuum of bosonic operators $\hat{\mathbf{b}}_\omega(\mathbf{r})$ that annihilate the environment ground state $|0_E\rangle$, and satisfy standard bosonic commutation relations:

$$\left[\hat{\mathbf{b}}_\omega(\mathbf{r}),\hat{\mathbf{b}}^\dagger_{\omega'}(\mathbf{r}')\right]=\delta(\omega-\omega')\delta(\mathbf{r}-\mathbf{r}')\mathbf{1}, \qquad \left[\hat{\mathbf{b}}_\omega(\mathbf{r}),\hat{\mathbf{b}}_{\omega'}(\mathbf{r}')\right]=0. \qquad (A1)$$

The environment Hamiltonian is

$$\hat{H}_{EM} = \int_0^\infty d\omega \int d^3\mathbf{r}\, \hbar\omega\, \hat{\mathbf{b}}_\omega^\dagger(\mathbf{r}) \cdot \hat{\mathbf{b}}_\omega(\mathbf{r}). \tag{A2}$$

The quantized electromagnetic fields $\hat{\mathbf{F}} = \begin{pmatrix} \hat{\mathbf{E}} & \hat{\mathbf{H}} \end{pmatrix}$ are written in terms of the operators $\hat{\mathbf{b}}_\omega$ and of the system Green's function $\overline{\mathcal{G}}(\mathbf{r},\mathbf{r}_0,\omega)$ as $\hat{\mathbf{F}} = \hat{\mathbf{F}}^- + \hat{\mathbf{F}}^+$ with:

$$\hat{\mathbf{F}}^-(\mathbf{r},t) = \int_{0^+}^\infty d\omega \int d^3\mathbf{r}' \sqrt{\frac{\hbar}{\pi|\omega|}} \overline{\mathcal{G}}(\mathbf{r},\mathbf{r}',\omega) \cdot \mathbf{R}_\omega(\mathbf{r}') \cdot \hat{\mathbf{b}}_\omega(\mathbf{r}') e^{-i\omega t}. \tag{A3}$$

The system Green's function $\overline{\mathcal{G}}(\mathbf{r},\mathbf{r}',\omega)$ (6×6 tensor) is the classical solution of the Maxwell's equation for a dipole-type excitation:

$$\hat{N} \cdot \overline{\mathcal{G}} - \omega \mathbf{M}(\mathbf{r},\omega) \cdot \overline{\mathcal{G}} = \omega \mathbf{1} \delta(\mathbf{r}-\mathbf{r}'), \quad \text{with } \hat{N} = \begin{pmatrix} 0 & i\nabla\times \\ -i\nabla\times & 0 \end{pmatrix}. \tag{A4}$$

In the above, $\hat{N}$ is a differential operator and $\mathbf{M} = \mathbf{M}(\mathbf{r},\omega)$ is a 6×6 material matrix that determines the environment response. It is written in terms of the permittivity and permeability tensors as:

$$\mathbf{M}(\mathbf{r},\omega) = \begin{pmatrix} \overline{\varepsilon}(\mathbf{r},\omega) & \mathbf{0}_{3\times 3} \\ \mathbf{0}_{3\times 3} & \mu_0 \mathbf{1}_{3\times 3} \end{pmatrix}. \tag{A5}$$

Furthermore, the matrix $\mathbf{R}_\omega$ in Eq. (A3) is defined by $\mathbf{R}_\omega = [\omega \mathbf{M}'']^{1/2}$ with $\mathbf{M}''(\omega) = (\mathbf{M} - \mathbf{M}^\dagger)/(2i)$.

For a real-valued $\omega$, the system Green's function satisfies the identity:

$$\overline{\mathcal{G}}(\mathbf{r},\mathbf{r}',\omega) - [\overline{\mathcal{G}}(\mathbf{r}',\mathbf{r},\omega)]^\dagger = \int d^3\mathbf{r}'' \overline{\mathcal{G}}(\mathbf{r},\mathbf{r}'',\omega) \cdot (\mathbf{M}(\mathbf{r}'',\omega) - \mathbf{M}^\dagger(\mathbf{r}'',\omega)) \cdot [\overline{\mathcal{G}}(\mathbf{r}',\mathbf{r}'',\omega)]^\dagger.$$

$$\tag{A6}$$

*B. Interaction with the environment and Born-Markov approximation*

We consider a multi-level atomic system described by the Hamiltonian $\hat{H}_{at}$. The coupling of the atom with the environment is modeled by the interaction Hamiltonian $\hat{H}_{int} = -\hat{\mathbf{F}}(\mathbf{r}_0) \cdot \hat{\mathbf{p}}$ where $\mathbf{r}_0$ stands for the position of the atom and $\hat{\mathbf{p}} = (\hat{\mathbf{p}}_e \quad \hat{\mathbf{p}}_m)^T$ is the (6-vector) transition dipole moment operator of the atom. The sub-component $\hat{\mathbf{p}}_e$ is the electric dipole moment operator, whereas $\hat{\mathbf{p}}_m = \mu_0 \hat{\mathbf{m}}_s$ is determined by the (spin) magnetic dipole moment operator ($\hat{\mathbf{m}}_s$).

We use the standard density matrix formalism to describe the time-evolution of the system state. The full-system density matrix is denoted by $\hat{\rho}$ and satisfies the von Neumann equation $\partial_t \hat{\rho} = \frac{i}{\hbar}[\hat{\rho}, \hat{H}]$ with $\hat{H} = \hat{H}_0 + \hat{H}_{int}$ and $\hat{H}_0 = \hat{H}_{at} + \hat{H}_{EM}$. Then, the interaction picture density matrix, $\hat{\rho}_I = e^{+i\frac{\hat{H}_0}{\hbar}t} \hat{\rho} e^{-i\frac{\hat{H}_0}{\hbar}t}$, satisfies:

$$\partial_t \hat{\rho}_I = \frac{i}{\hbar}\left[\hat{\rho}_I, \hat{H}_{int}(t)\right], \tag{A7}$$

with $\hat{H}_{int}(t) = e^{i\frac{\hat{H}_0}{\hbar}t} \hat{H}_{int} e^{-i\frac{\hat{H}_0}{\hbar}t}$.

In the Born approximation, the density matrix is assumed to satisfy $\hat{\rho}_I(t) \approx \hat{\rho}_{S,I}(t) \otimes \hat{\rho}_E$ at all time instants, where $\hat{\rho}_{S,I}$ and $\hat{\rho}_E$ depend only on the atomic (environment) states, respectively. Typically, the environment density matrix $\hat{\rho}_E$ is supposed to be time-invariant. In this study, we consider the influence of the intrinsic spin magnetic moment of an electron on the macroscopic gyrotropic response of the environment. Thus, we take into account that the static-type magnetic field $\mathbf{B}_0$ arising from the electron spin can modify the permittivity of the plasmonic particle, consequently affecting the environment ground-state $|0_E\rangle$. It is underlined that the value of $\mathbf{B}_0$ depends on the atomic configuration and governs the macroscopic permittivity response of the nanoparticle.

We will consider that $\hat{\rho}_E \approx |0_E\rangle\langle 0_E|$ with $|0_E\rangle$ the (parametric) ground state of the environment, which as explained above depends on $\mathbf{B}_0$. Then, proceeding in a standard manner, one finds that the dynamics of the reduced density matrix $\hat{\rho}_{S,I}(t)$ is determined by

$$\partial_t \hat{\rho}_{S,I} = -\frac{1}{\hbar^2}\int_0^{+\infty} \mathrm{tr}_E\left\{\left[\left[\hat{\rho}_I(t-\tau), \hat{H}_{int}(t-\tau)\right], \hat{H}_{int}(t)\right]\right\}d\tau. \tag{A8}$$

The operator $\mathrm{tr}_E\{...\}$ represents the trace over the environment degrees of freedom. It is supposed that at initial time $\mathrm{tr}_E\left\{\left[\hat{\rho}_I(t_0), \hat{H}_{int}(t)\right]\right\} = 0$.

Next, we use the Markov approximation so that $\hat{\rho}_I(t-\tau) \to \hat{\rho}_I(t)$ in Eq. (A8). This leads to:

$$\begin{aligned}\partial_t \hat{\rho}_{S,I} = &-\frac{1}{\hbar^2}\int_0^{+\infty} \mathrm{tr}_E\left\{\hat{\rho}_I(t)\hat{H}_{int}(t-\tau)\hat{H}_{int}(t)\right\}d\tau \\ &-\frac{1}{\hbar^2}\int_0^{+\infty} \mathrm{tr}_E\left\{-\hat{H}_{int}(t-\tau)\hat{\rho}_I(t)\hat{H}_{int}(t)\right\}d\tau \\ &-\frac{1}{\hbar^2}\int_0^{+\infty} \mathrm{tr}_E\left\{-\hat{H}_{int}(t)\hat{\rho}_I(t)\hat{H}_{int}(t-\tau)\right\}d\tau \\ &-\frac{1}{\hbar^2}\int_0^{+\infty} \mathrm{tr}_E\left\{\hat{H}_{int}(t)\hat{H}_{int}(t-\tau)\hat{\rho}_I(t)\right\}d\tau\end{aligned} \tag{A9}$$

### C. Derivation of the master equation

In order to simplify Eq. (A9), first we use Eq. (A6) and $\hat{\rho}_E \approx |0_E\rangle\langle 0_E|$ to obtain:

$$\mathrm{tr}_E\left\{\hat{\rho}_E\hat{\mathbf{F}}(\mathbf{r},t)\otimes\hat{\mathbf{F}}(\mathbf{r}',t-\tau)\right\}=\frac{\hbar}{2\pi i}\int_0^\infty d\omega e^{-i\omega\tau}\left[\overline{\mathcal{G}}(\mathbf{r},\mathbf{r}',\omega)-\left[\overline{\mathcal{G}}(\mathbf{r}',\mathbf{r},\omega)\right]^\dagger\right],\quad\text{(A10a)}$$

$$\mathrm{tr}_E\left\{\hat{\mathbf{F}}(\mathbf{r},t)\hat{\rho}_E\otimes\hat{\mathbf{F}}(\mathbf{r}',t-\tau)\right\}=-\frac{\hbar}{2\pi i}\int_0^\infty d\omega e^{+i\omega\tau}\left[\overline{\mathcal{G}}(\mathbf{r},\mathbf{r}',\omega)-\left[\overline{\mathcal{G}}(\mathbf{r}',\mathbf{r},\omega)\right]^\dagger\right]^*.\quad\text{(A10b)}$$

Next, we substitute $\hat{H}_{\mathrm{int}}(t)=-\hat{\mathbf{F}}(\mathbf{r}_0,t)\cdot\hat{\mathbf{p}}(t)$ into Eq. (A9), taking into account the above identities. After some simplifications, using $\hat{\mathbf{F}}=\hat{\mathbf{F}}^-+\hat{\mathbf{F}}^+$ with $\hat{\mathbf{F}}^-$ given by Eq. (A3), one may show that the system dynamics is controlled by $\partial_t\hat{\rho}_{\mathrm{S,I}}=\mathcal{L}\hat{\rho}_{\mathrm{S,I}}$ with the Lindblad operator $\mathcal{L}$ given by:

$$\begin{aligned}\mathcal{L}\hat{\rho}_{\mathrm{S,I}}=\\-\frac{1}{2\pi\hbar i}\hat{\rho}_{\mathrm{S,I}}(t)\int_0^{+\infty}d\omega\left(\int_0^{+\infty}\hat{\mathbf{p}}(t-\tau)e^{+i\omega\tau}d\tau\right)\mathbf{A}(\omega)\cdot\hat{\mathbf{p}}(t)\\-\frac{1}{2\pi\hbar i}\hat{\mathbf{p}}(t)\cdot\int_0^{+\infty}d\omega\mathbf{A}(\omega)\cdot\left(\int_0^{+\infty}e^{-i\omega\tau}\hat{\mathbf{p}}(t-\tau)d\tau\right)\hat{\rho}_{\mathrm{S,I}}(t)\\-\frac{1}{2\pi\hbar i}\int_0^{+\infty}d\omega\left(\int_0^{+\infty}\hat{\mathbf{p}}(t-\tau)e^{-i\omega\tau}d\tau\right)\cdot\mathbf{A}^*(\omega)\hat{\rho}_{\mathrm{S,I}}(t)\cdot\hat{\mathbf{p}}(t)\\-\frac{1}{2\pi\hbar i}\hat{\mathbf{p}}(t)\cdot\int_0^{+\infty}d\omega\mathbf{A}^*(\omega)\hat{\rho}_{\mathrm{S,I}}(t)\cdot\left(\int_0^{+\infty}e^{+i\omega\tau}\hat{\mathbf{p}}(t-\tau)d\tau\right)\end{aligned}\quad\text{(A11)}$$

and $\mathbf{A}(\omega)=\overline{\mathcal{G}}(\mathbf{r}_0,\mathbf{r}_0,\omega)-\left[\overline{\mathcal{G}}(\mathbf{r}_0,\mathbf{r}_0,\omega)\right]^\dagger$. It is underlined that the master equation $\partial_t\hat{\rho}_{\mathrm{S,I}}=\mathcal{L}\hat{\rho}_{\mathrm{S,I}}$ applies to an arbitrary atomic system, i.e., may include the contributions from an arbitrary number of energy levels and can also account for the effect of the spin magnetic moment.

### D. Degenerate two-level systems

Next, we consider the particular case of an atom with only two (degenerate) energy levels. The ground states are denoted by $|g_i\rangle$ and the excited states by $|e_j\rangle$, with $i=1,\ldots,$ and $j=1,\ldots$, running over the different degenerate states. The energy difference between the ground and excited states is $\hbar\omega_a$ so that the atom Hamiltonian is $\hat{H}_{at} = \hbar\omega_a \sum_j |e_j\rangle\langle e_j|$. Furthermore, we suppose the environment does not have a magnetic response ($\mu = \mu_0$). In such a case, the effects of the spin magnetic moment operator are expected to be negligible and we can use $\hat{\mathbf{p}} \approx (\hat{\mathbf{p}}_e \quad \mathbf{0})^T$. In fact, the interactions with the environment are mainly ruled by quasi-static interactions (see Appendix C) and an environment with $\mu = \mu_0$ does not back-scatter the magnetic fields created by the magnetic moment $\hat{\mathbf{p}}_m = \mu_0 \hat{\mathbf{m}}_s$. It is however important to underline the effects of $\hat{\mathbf{m}}_s$ are not dropped altogether in our analysis as they influence the gyroelectric response of the environment. The effects of radiation loss due to the precession of the spin magnetic moment are discussed in Appendix E.

Within these hypotheses, the dipole operator may be assumed of the type $\hat{\mathbf{p}}(t) = \hat{\mathbf{p}}^-(t) + \hat{\mathbf{p}}^+(t)$ with $\hat{\mathbf{p}}^-(t) = \hat{\mathbf{p}}^- e^{-i\omega_a t}$, $\hat{\mathbf{p}}^+(t) = \hat{\mathbf{p}}^+ e^{+i\omega_a^* t}$ and $\hat{\mathbf{p}}^\pm$ ($\hat{\mathbf{p}}^+ = [\hat{\mathbf{p}}^-]^\dagger$) the operators in the Schrödinger picture determined only by the electric dipole operator ($\hat{\mathbf{p}} \approx (\hat{\mathbf{p}}_e \quad \mathbf{0})^T$). For convenience, we include in $\omega_a$ a small positive imaginary part ($\omega_a \to \omega_a + 0^+ i$) that accounts for a slow (adiabatic) switching of the interaction term. Typically, $\hat{\mathbf{p}}^-$ is an operator of the type $\hat{\mathbf{p}}^- = \sum_{i,j} \boldsymbol{\gamma}_{ij} |g_i\rangle\langle e_j|$, where $\boldsymbol{\gamma}_{ij}$ is the transition electric dipole moment (6 vector with the last 3 entries equal to zero) of the electronic transition $|e_j\rangle \to |g_i\rangle$.

In the previous conditions,

$$\int_0^{+\infty} e^{-i\omega\tau}\hat{\mathbf{p}}(t-\tau)d\tau = \hat{\mathbf{p}}^-(t)\frac{1}{i(\omega-\omega_a)} + \hat{\mathbf{p}}^+(t)\frac{1}{i(\omega+\omega_a^*)}. \tag{A12}$$

Then, after straightforward simplifications Eq. (A11) becomes:

$$\begin{aligned}\mathcal{L}\hat{\rho}_{S,I} =& \\ &-\frac{1}{2\pi\hbar}\hat{\rho}_{S,I}(t)\int_0^\infty d\omega\left(\hat{\mathbf{p}}^-(t)\cdot\frac{\mathbf{A}(\omega)}{(\omega+\omega_a)} + \hat{\mathbf{p}}^+(t)\cdot\frac{\mathbf{A}(\omega)}{(\omega-\omega_a^*)}\right)\cdot\hat{\mathbf{p}}(t)\\ &\frac{1}{2\pi\hbar}\hat{\mathbf{p}}(t)\cdot\int_0^\infty d\omega\left(\frac{\mathbf{A}(\omega)}{(\omega-\omega_a)}\cdot\hat{\mathbf{p}}^-(t) + \frac{\mathbf{A}(\omega)}{(\omega+\omega_a^*)}\cdot\hat{\mathbf{p}}^+(t)\right)\hat{\rho}_{S,I}(t)\\ &+\frac{1}{2\pi\hbar}\int_0^\infty d\omega\left(\hat{\mathbf{p}}^-(t)\cdot\frac{\mathbf{A}^*(\omega)}{(\omega-\omega_a)} + \hat{\mathbf{p}}^+(t)\cdot\frac{\mathbf{A}^*(\omega)}{(\omega+\omega_a^*)}\right)\cdot\hat{\rho}_{S,I}(t)\cdot\hat{\mathbf{p}}(t)\\ &-\frac{1}{2\pi\hbar}\hat{\mathbf{p}}(t)\cdot\hat{\rho}_{S,I}(t)\int_0^\infty d\omega\left(\frac{\mathbf{A}^*(\omega)}{(\omega+\omega_a)}\cdot\hat{\mathbf{p}}^-(t) + \frac{\mathbf{A}^*(\omega)}{(\omega-\omega_a^*)}\hat{\mathbf{p}}^+(t)\right)\end{aligned} \tag{A13}$$

Next, we switch back to the Schrödinger picture, so that the master equation becomes:

$$\partial_t \hat{\rho}_S = -i\frac{1}{\hbar}\left[\hat{H}_{at}, \hat{\rho}_S\right] + \mathcal{L}\hat{\rho}_S, \tag{A14}$$

where $\hat{H}_{at}$ is the atom Hamiltonian, $\hat{\rho}_S = e^{-i\frac{\hat{H}_{at}}{\hbar}t}\hat{\rho}_{S,I}e^{-i\frac{\hat{H}_{at}}{\hbar}t}$ is the reduced density matrix in the Schrödinger picture. Noting that $\mathbf{A}(\omega)/2i$ is a Hermitian matrix, one may show that the Lindblad operator in the Schrödinger picture satisfies:

$$\begin{aligned}\mathcal{L}\hat{\rho}_S = &-\frac{i}{\hbar}\hat{\rho}_S(t)\left(\hat{\mathbf{p}}^-\cdot\left[\overline{\mathcal{G}}^-\right]^*\cdot\hat{\mathbf{p}} + \hat{\mathbf{p}}^+\cdot\left[\overline{\mathcal{G}}^+\right]^\dagger\cdot\hat{\mathbf{p}}\right)\\ &+\frac{i}{\hbar}\left(\hat{\mathbf{p}}\cdot\overline{\mathcal{G}}^+\cdot\hat{\mathbf{p}}^- + \hat{\mathbf{p}}\cdot\left[\overline{\mathcal{G}}^-\right]^T\cdot\hat{\mathbf{p}}^+\right)\hat{\rho}_S(t)\\ &\frac{-i}{\hbar}\left(\hat{\mathbf{p}}^-\cdot\left[\overline{\mathcal{G}}^+\right]^T\hat{\rho}_S(t)\cdot\hat{\mathbf{p}} + \hat{\mathbf{p}}^+\cdot\overline{\mathcal{G}}^-\hat{\rho}_S(t)\cdot\hat{\mathbf{p}}\right)\\ &+\frac{i}{\hbar}\left(\hat{\mathbf{p}}\cdot\left[\overline{\mathcal{G}}^-\right]^\dagger\hat{\rho}_S(t)\cdot\hat{\mathbf{p}}^- + \hat{\mathbf{p}}\cdot\left[\overline{\mathcal{G}}^+\right]^*\hat{\rho}_S(t)\cdot\hat{\mathbf{p}}^+\right)\end{aligned} \tag{A15}$$

where we introduced the tensors:

$$\overline{\mathcal{G}}^+ = \frac{1}{2\pi i}\int_{0^+}^{\infty} d\omega \frac{\mathbf{A}(\omega)}{(\omega-\omega_a)}, \qquad \overline{\mathcal{G}}^- = \frac{-1}{2\pi i}\int_{0^+}^{\infty} d\omega \frac{\mathbf{A}^*(\omega)}{(\omega+\omega_a^*)}. \tag{A16}$$

In the above, the symbols † and $T$ represent the Hermitian conjugate matrix and the transpose matrix, respectively. Note that in the Schrödinger picture the operators $\hat{\mathbf{p}}^\pm$ are time independent. Furthermore, it is worth pointing out that the rotating wave approximation was not applied.

In the calculation of $\overline{\mathcal{G}}^-$ the frequency $\omega_a$ can be assumed real-valued because the integral is free of singularities. Consequently, $\overline{\mathcal{G}}^-$ is a Hermitian tensor:

$$\overline{\mathcal{G}}^- = \left[\overline{\mathcal{G}}^-\right]^\dagger. \tag{A17}$$

In contrast, the integrand associated with $\overline{\mathcal{G}}^+$ has a singularity and hence the imaginary part of $\omega_a$ cannot be dropped. In the next subsection, it is shown that $\overline{\mathcal{G}}^+ = \overline{\mathcal{G}} - \overline{\mathcal{G}}^-$ with $\overline{\mathcal{G}} = \overline{\mathcal{G}}(\mathbf{r}_0, \mathbf{r}_0, \omega_a)$ the system Green's function. Furthermore, we shall show that $\overline{\mathcal{G}}^-$ can be written in terms of an integral over the imaginary frequency axis.

Using $\overline{\mathcal{G}}^- = \left[\overline{\mathcal{G}}^-\right]^\dagger$ in Eq. (A15), one can prove that for the case of a non-degenerate two-level system (with a single excited state and single ground state) it reduces to the results derived in previous works [55, 56, 57] when the rotating wave approximation is applied to the Lindblad operator.

*E. The functions $\overline{\mathcal{G}}^\pm$*

Let us now relate the tensors $\overline{\mathcal{G}}^\pm$ (see also Ref. [58]). To begin with we use $\mathbf{A}(\omega) = \overline{\mathcal{G}}(\mathbf{r}_0,\mathbf{r}_0,\omega) - \left[\overline{\mathcal{G}}(\mathbf{r}_0,\mathbf{r}_0,\omega)\right]^\dagger$ to write:

$$\overline{\mathcal{G}}^- = \left[\frac{1}{2\pi i}\int_{0^+}^{\infty} d\omega \frac{\overline{\mathcal{G}}(\mathbf{r}_0,\mathbf{r}_0,\omega)}{(\omega+\omega_a)}\right]^* + \frac{1}{2\pi i}\int_{0^+}^{\infty} d\omega \frac{\left[\overline{\mathcal{G}}(\mathbf{r}_0,\mathbf{r}_0,\omega)\right]^T}{(\omega+\omega_a^*)}. \tag{A18}$$

Performing a Wick rotation, one can deform the integration path to the positive imaginary frequency axis without crossing any poles of the integrand in the 1st quadrant of the complex plane ($\text{Re}\{\omega\} > 0$ and $\text{Im}\{\omega\} > 0$):

$$\overline{\mathcal{G}}^- = \left[\frac{1}{2\pi}\int_{0^+}^{\infty} d\xi \frac{\overline{\mathcal{G}}(\mathbf{r}_0,\mathbf{r}_0,i\xi)}{(i\xi+\omega_a)}\right]^* + \frac{1}{2\pi}\int_{0^+}^{\infty} d\xi \frac{\left[\overline{\mathcal{G}}(\mathbf{r}_0,\mathbf{r}_0,i\xi)\right]^T}{(i\xi+\omega_a^*)}. \tag{A19}$$

Taking into account that $\overline{\mathcal{G}}(\mathbf{r}_0,\mathbf{r}_0,i\xi)$ is real-valued, it follows that:

$$\overline{\mathcal{G}}^- = \frac{1}{2\pi}\int_{0^+}^{\infty} d\xi \left[\frac{\overline{\mathcal{G}}(\mathbf{r}_0,\mathbf{r}_0,i\xi)}{(\omega_a-i\xi)} + \frac{\left[\overline{\mathcal{G}}(\mathbf{r}_0,\mathbf{r}_0,i\xi)\right]^T}{(\omega_a+i\xi)}\right], \tag{A20}$$

where we dropped the infinitesimal imaginary part ($0^+i$) of $\omega_a$ in the final formula.

We can proceed in a similar manner for the tensor $\overline{\mathcal{G}}^+$. Different from the previous case, now the Wick rotation will cross a pole of the integrand at $\omega = \omega_a$. A straightforward analysis shows that:

$$\overline{\mathcal{G}}^+ = \overline{\mathcal{G}}(\mathbf{r}_0,\mathbf{r}_0,\omega_a) - \overline{\mathcal{G}}^-. \tag{A21}$$

The Green's function $\overline{\mathcal{G}}(\mathbf{r}_0,\mathbf{r}_0,\omega_a)$ can be decomposed as $\overline{\mathcal{G}} = \overline{\mathcal{G}}^{\text{self}} + \overline{\mathcal{G}}^{\text{scat}}$. The first term is the divergent self-contribution (the real part of $\overline{\mathcal{G}}^{\text{self}}$ has a singularity), which is discarded. The second term is the scattering part of the Green's function.

## Appendix B: Master equation for the ground-subspace

In this Appendix, we focus our analysis in the ground-space of the reduced density matrix. The ground subspace is generated by objects of the type $|g_m\rangle\langle g_n|$ with $m,n$ running over the degenerate ground states. As in subsection D of Appendix A, we suppose that the dipole moment operator is $\hat{\mathbf{p}} = \hat{\mathbf{p}}^- + \hat{\mathbf{p}}^+$ with $\hat{\mathbf{p}}^- = \sum_{i,j} \gamma_{ij} |g_i\rangle\langle e_j|$ with $\hat{\mathbf{p}} \approx (\hat{\mathbf{p}}_e \quad \mathbf{0})^T$.

To begin with, let us prove that the ground-space is closed, i.e., that when the Lindblad operator [Eq. (A15)] acts on an element of the ground-subspace it generates another element of the ground-subspace. To this end, we note that if $\hat{\rho}_S(t)$ is in the ground-subspace then $\hat{\rho}_S(t)\hat{\mathbf{p}}^+ = \hat{\mathbf{p}}^- \hat{\rho}_S(t) = 0$. Taking also into account that $\overline{\mathcal{G}}^- = \left[\overline{\mathcal{G}}^-\right]^\dagger$ [Eq. (A17)] it follows that

$$\mathcal{L}\hat{\rho}_S\big|_{\text{ground}} = \frac{i}{\hbar}\left(\hat{\mathbf{p}} \cdot \left[\overline{\mathcal{G}}^-\right]^T \cdot \hat{\mathbf{p}}^+ \hat{\rho}_S(t) - \hat{\rho}_S(t)\hat{\mathbf{p}}^- \cdot \left[\overline{\mathcal{G}}^-\right]^T \cdot \hat{\mathbf{p}}\right). \tag{B1}$$

Furthermore, it is clear that if $\hat{\rho}_S(t)$ is in the ground-subspace then $\hat{p}_i^+ \hat{p}_{i'}^+ \hat{\rho}_S(t) = 0 = \hat{\rho}_S(t)\hat{p}_i^- \hat{p}_{i'}^-$. This observation shows that:

$$\mathcal{L}\hat{\rho}_S\big|_{\text{ground}} = \frac{i}{\hbar}\left(\hat{\mathbf{p}}^- \cdot \left[\overline{\mathcal{G}}^-\right]^T \cdot \hat{\mathbf{p}}^+ \hat{\rho}_S(t) - \hat{\rho}_S(t)\hat{\mathbf{p}}^- \cdot \left[\overline{\mathcal{G}}^-\right]^T \cdot \hat{\mathbf{p}}^+\right). \tag{B2}$$

Evidently, the element in the right-hand side of the above equation is an element of the ground-subspace. This confirms that the ground-subspace is indeed closed.

Equation (B2) can be rewritten in a more compact way in terms of a commutator:

$$\mathcal{L}\hat{\rho}_S\big|_{\text{ground}} = \frac{i}{\hbar}\left[\hat{\mathbf{p}}^- \cdot \left[\overline{\mathcal{G}}^-\right]^T \cdot \hat{\mathbf{p}}^+, \hat{\rho}_S(t)\right]. \tag{B3}$$

Thereby, it follows that the dynamics of the reduced density matrix $\partial_t \hat{\rho}_S = -i\frac{1}{\hbar}\left[\hat{H}_{at}, \hat{\rho}_S\right] + \mathcal{L}\hat{\rho}_S$ [Eq. (A14)] may reduced to a von Neumann equation of the type:

$$\partial_t \hat{\rho}_S = -i\frac{1}{\hbar}\left[\hat{H}_{ef}, \hat{\rho}_S\right], \qquad \text{(ground subspace)}, \tag{B4}$$

with $\hat{H}_{ef} = \hat{H}_{at} - \hat{\mathbf{p}}^- \cdot \left[\overline{\mathcal{G}}^-\right]^T \cdot \hat{\mathbf{p}}^+$ the effective Hamiltonian of the ground. Since $\hat{H}_{at}$ restricted to the ground subspace is a constant, one can take:

$$\hat{H}_{ef} = -\hat{\mathbf{p}}^- \cdot \left[\overline{\mathcal{G}}^-\right]^T \cdot \hat{\mathbf{p}}^+, \qquad \text{(ground subspace)}. \tag{B5}$$

The effective Hamiltonian is determined by the Green's function $\overline{\mathcal{G}}^-$ [Eq. (A20)], which depends on the static magnetic field $\mathbf{B}_0$ created by the intrinsic spin magnetic moment of the electron. Thus, as further discussed in the main text, $\hat{H}_{ef}$ is a function of the reduced density matrix and Eq. (B4) is a nonlinear master equation. As the reduced density matrix is ruled by a von Neumann equation, for a pure state it may be assumed of the form $\hat{\rho}_S(t) = |\psi(t)\rangle\langle\psi(t)|$ for all time instants, with $|\psi(t)\rangle$ a vector in the ground subspace satisfying the nonlinear Schrödinger equation:

$$i\hbar\frac{\partial}{\partial t}|\psi(t)\rangle = \hat{H}_{ef}(\mathbf{B}_0)|\psi(t)\rangle. \tag{B6}$$

For the geometry of the main text, the static magnetic field is given by:

$$\mathbf{B}_0 = \frac{\mu_0}{4\pi d^3}(3\hat{\mathbf{z}}\otimes\hat{\mathbf{z}}-\mathbf{1})\cdot\mathbf{m}_s, \qquad \text{with}\quad \mathbf{m}_s = \langle\psi(t)|\hat{\mathbf{m}}_s|\psi(t)\rangle \tag{B7}$$

and $\hat{\mathbf{m}}_s$ the electron spin magnetic moment operator. It is underlined that the previous analysis applies to an arbitrary degenerate two-level system with an arbitrary number of degenerate excited (ground) states.

Supposing that the ground subspace is generated by two states (e.g., connected by time-reversal symmetry), $|g_1\rangle$ and $|g_2\rangle$ it is possible to write $|\psi(t)\rangle = c_1(t)|g_1\rangle + c_2(t)|g_2\rangle$. The dynamics of the coefficients $c_1(t), c_2(t)$ is ruled by:

$$i\hbar\partial_t \begin{pmatrix} c_1 \\ c_2 \end{pmatrix} = \mathbf{H}_{\text{ef}}(c_1,c_2)\cdot\begin{pmatrix} c_1 \\ c_2 \end{pmatrix}, \tag{B8}$$

where $\mathbf{H}_{\text{ef}}(c_1,c_2) = [h_{m,n}]$ is a 2×2 matrix ($m,n=1,2$) with entries

$$h_{m,n} = \langle g_m|\hat{H}_{\text{ef}}(\mathbf{B}_0)|g_n\rangle. \tag{B9}$$

Using Eq. (A20), it is simple to show that:

$$\begin{aligned}h_{m,n} = &-\frac{1}{2\pi}\int_{0^+}^{\infty} d\xi \langle g_m|\hat{\mathbf{p}}^-\cdot[\overline{\mathcal{G}}(\mathbf{r}_0,\mathbf{r}_0,i\xi)]^T\cdot\hat{\mathbf{p}}^+|g_n\rangle \frac{1}{\omega_a - i\xi} \\ &-\frac{1}{2\pi}\int_{0^+}^{\infty} d\xi \langle g_m|\hat{\mathbf{p}}^-\cdot\overline{\mathcal{G}}(\mathbf{r}_0,\mathbf{r}_0,i\xi)\cdot\hat{\mathbf{p}}^+|g_n\rangle \frac{1}{\omega_a + i\xi}\end{aligned} \tag{B10}$$

This can also be rewritten as:

$$h_{m,n}(\mathbf{B}_0) = -\text{tr}\left\{\langle g_m|\hat{\mathbf{p}}^-\otimes\hat{\mathbf{p}}^+|g_n\rangle\cdot\left(\frac{1}{2\pi}\int_{0^+}^{\infty}d\xi\frac{\overline{\mathcal{G}}(\mathbf{r}_0,\mathbf{r}_0,i\xi)}{\omega_a-i\xi}+\frac{[\overline{\mathcal{G}}(\mathbf{r}_0,\mathbf{r}_0,i\xi)]^T}{\omega_a+i\xi}\right)\right\}. \tag{B11}$$

It should be noted that in the above formulas the Green's function depends on $\mathbf{B}_0$, and hence also on the coefficients $c_1(t), c_2(t)$. It may be checked that the effective

Hamiltonian precisely agrees with the result obtained by applying lowest-order perturbation theory to the full Hamiltonian (see Appendix E of Ref. [44]). As previously noted, the above results apply to a degenerate two-level system with the dimension of the excited subspace arbitrary. In the main text, we restrict our attention to the case where the excited subspace is generated by only two states ($|e_1\rangle$ and $|e_2\rangle$) linked by time-reversal symmetry. For completeness, we mention that had the contribution from the magnetic dipole been retained, Eq. (B11) would get an additional term with the same structure with $\hat{\mathbf{p}}^- = \hat{\mathbf{p}}^+ \rightarrow \begin{pmatrix} 0 & \hat{\mathbf{p}}_m \end{pmatrix}^T$ and $\omega_a = 0$.

**Appendix C: Effective Hamitonian with a quasi-static approximation**

In the following, we obtain approximate analytical formulas for the effective Hamiltonian matrix elements [Eq. (8)] neglecting retardation effects (quasi-static approximation).

The elements $h_{m,n}$ are written in terms of the system Green's function $\overline{\mathcal{G}}(\mathbf{r}_0,\mathbf{r}_0,i\xi)$ evaluated in the imaginary frequency axis. The Green's function can be decomposed into self and scattering terms: $\overline{\mathcal{G}} = \overline{\mathcal{G}}^{\text{self}} + \overline{\mathcal{G}}^{\text{scat}}$. The self-term corresponds to the field radiated by a dipole (electric or magnetic) alone in free-space [see Eq. (A4)]. We will evaluate $\overline{\mathcal{G}}$ using a quasi-static approximation, which is typically very accurate for near-field interactions. In the quasi-static limit and for a nanoparticle with a trivial magnetic response ($\mu = \mu_0$) the magnetic part of $\overline{\mathcal{G}}^{\text{scat}}$ vanishes. This is so because the magnetic particle does not scatter magnetic fields when $\mu = \mu_0$. Thus, we can safely focus on the electric part of $\overline{\mathcal{G}}^{\text{scat}}$ associated with electric excitations (see also Appendix A, Sect. *D*).

The quasi-static field created by an electric dipole with dipole moment $\mathbf{p}_e$ is $\mathbf{E}^{\text{self}} = \frac{1}{4\pi\varepsilon_0 r^3}(3\hat{\mathbf{r}} \otimes \hat{\mathbf{r}} - \mathbf{1}) \cdot \mathbf{p}_e$. The Green's function term $\overline{\mathcal{G}}^{\text{scat}}$ gives the field back-scattered by the nanoparticle due to $\mathbf{E}^{\text{self}}$. The local (incident) field at the nanoparticle position is:

$$\mathbf{E}_{\text{np}} = \mathbf{C}^{\text{int}} \cdot \frac{\mathbf{p}_e}{\varepsilon_0}, \quad \text{with} \quad \mathbf{C}^{\text{int}} = \frac{1}{4\pi d^3}(3\hat{\mathbf{z}} \otimes \hat{\mathbf{z}} - \mathbf{1}). \tag{C1}$$

We took into account that the relative position of the atom with respect to the nanoparticle is given by the vector $d\,\hat{\mathbf{z}}$ (see Fig. 1). The field $\mathbf{E}_{\text{np}}$ separates the electrical charges of the nanoparticle creating an electric dipole $\mathbf{p}_{\text{np}}$. It is given by $\mathbf{p}_{\text{np}} = \varepsilon_0 \boldsymbol{\alpha}_{\text{np}} \cdot \mathbf{E}_{\text{np}}$, where $\boldsymbol{\alpha}_{\text{np}}(\omega)$ is the nanoparticle polarizability. The electric polarizability of the gyroelectric nanosphere is calculated in Appendix D using a quasi-static approximation [Eq. (D8)].

The field back-scattered at the atom position is evidently $\mathbf{E}^{\text{bs}} = \mathbf{C}^{\text{int}} \cdot \mathbf{p}_{\text{np}}/\varepsilon_0$. It can also be written in terms of the primary excitation as $\mathbf{E}^{\text{bs}} = \mathbf{C}^{\text{int}} \cdot \boldsymbol{\alpha}_{\text{np}} \cdot \mathbf{C}^{\text{int}} \cdot \mathbf{p}_e/\varepsilon_0$. This discussion proves that the scattering part of the Green's function is given by the tensor:

$$\overline{\mathcal{G}}^{\text{scat}}(\mathbf{r}_0, \mathbf{r}_0, i\xi) = \frac{1}{\varepsilon_0} \mathbf{C}^{\text{int}} \cdot \boldsymbol{\alpha}_{\text{np}}(i\xi) \cdot \mathbf{C}^{\text{int}}. \tag{C2}$$

In summary, within a quasi-static approximation the integral in Eq. (8) can be evaluated with the substitution $\overline{\mathcal{G}} \to \overline{\mathcal{G}}^{\text{scat}}$ with $\overline{\mathcal{G}}^{\text{scat}}$ defined as in Eq. (C2). Again we underline that even though the atom has a spin magnetic dipole moment, the magnetic part of the Green's function can be safely ignored because the nanoparticle itself is assumed nonmagnetic ($\mu = \mu_0$). Nevertheless, the effects of the spin magnetic moment are still relevant as they tailor the electric response of the nanoparticle.

Next, we obtain an explicit expression for Eq. (8). To begin we note that $\mathbf{Q}^{(m,n)} = \langle g_m | \hat{\mathbf{p}}^- \otimes \hat{\mathbf{p}}^+ | g_n \rangle$ has the following symmetries:

$$\mathbf{Q}^{(m,n)} = \mathbf{Q}^{(n,m)\dagger} \quad \text{and} \quad \mathbf{Q}^{(m,n)} = \langle \mathcal{T} g_m | \hat{\mathbf{p}}^- \otimes \hat{\mathbf{p}}^+ | \mathcal{T} g_n \rangle^*. \tag{C3}$$

The first identity follows from the fact that $\hat{\mathbf{p}}$ is Hermitian and $\mathbf{Q}^{(m,n)}$ can be expressed as $\mathbf{Q}^{(m,n)} = \langle g_m | \hat{\mathbf{p}} \otimes \hat{\mathbf{p}} | g_n \rangle$. On the other hand, the second identity uses the formula $\langle g_m | \hat{A} | g_n \rangle = \langle \mathcal{T} g_m | \hat{A} | \mathcal{T} g_n \rangle^*$, which is valid for a generic time-reversal invariant operator $\hat{A}$ [44]. In particular, combining the two identities and using $|g_2\rangle = |\mathcal{T} g_1\rangle$ and $|g_1\rangle = -|\mathcal{T} g_2\rangle$ one sees that the $\mathbf{Q}^{(m,n)}$ matrices are anti-symmetric for $m \neq n$ and that $\mathbf{Q}^{(1,1)} = \mathbf{Q}^{(2,2),*}$. The discussed symmetries are in agreement with Eq. (11) for a Kramers two-level system.

On the other hand, the Green's function can be decomposed into symmetric and anti-symmetric parts $\overline{\mathcal{G}} = \overline{\mathcal{G}}_S + \overline{\mathcal{G}}_{AS}$ with $\overline{\mathcal{G}}_S = \frac{1}{2}\left(\overline{\mathcal{G}} + \overline{\mathcal{G}}^T\right)$ and $\overline{\mathcal{G}}_{AS} = \frac{1}{2}\left(\overline{\mathcal{G}} - \overline{\mathcal{G}}^T\right)$. Since the trace of the product of a symmetric and anti-symmetric matrix vanishes, it follows that Eq. (8) can be written as:

$$\begin{aligned} h_{m,n}(\mathbf{B}_0) = &-\delta_{mn} \frac{1}{2\pi} \int_{0^+}^{\infty} d\xi \, \text{tr}\left\{\mathbf{Q}^{(m,m)} \cdot \overline{\mathcal{G}}_S\right\} \left(\frac{1}{\omega_a - i\xi} + \frac{1}{\omega_a + i\xi}\right) \\ &- \frac{1}{2\pi} \int_{0^+}^{\infty} d\xi \, \text{tr}\left\{\mathbf{Q}^{(m,n)} \cdot \overline{\mathcal{G}}_{AS}\right\} \left(\frac{1}{\omega_a - i\xi} - \frac{1}{\omega_a + i\xi}\right). \end{aligned} \tag{C4}$$

Using now $\mathbf{Q}^{(1,1)} = \mathbf{Q}^{(2,2),*}$ and $\mathbf{Q}^{(m,m)} = \mathbf{Q}^{(m,m)\dagger}$ one sees that $\text{tr}\{\mathbf{Q}^{(1,1)} \cdot \overline{\mathcal{G}}_S\} = \text{tr}\{\mathbf{Q}^{(2,2)} \cdot \overline{\mathcal{G}}_S\}$. It was taken into account that for generic matrices $\mathbf{A}, \mathbf{B}$ one has $\text{tr}\{\mathbf{A} \cdot \mathbf{B}\} = \text{tr}\{\mathbf{B} \cdot \mathbf{A}\} = \text{tr}\{\mathbf{B}^T \cdot \mathbf{A}^T\}$. Thereby, we find that:

$$h_{m,n}(\mathbf{B}_0) = \delta_{mn}\mathcal{E}_0(\mathbf{B}_0) - \frac{1}{2\pi}\int_{0^+}^{\infty} d\xi \, \text{tr}\{\mathbf{Q}^{(m,n)} \cdot \overline{\mathcal{G}}_{AS}\} \frac{2i\xi}{\omega_a^2 + \xi^2}, \quad \text{with} \tag{C5a}$$

$$\mathcal{E}_0(\mathbf{B}_0) = -\frac{1}{2\pi}\int_{0^+}^{\infty} d\xi \, \text{tr}\{\mathbf{Q}^{(m,m)} \cdot \overline{\mathcal{G}}_S\} \frac{2\omega_a}{\omega_a^2 + \xi^2}. \tag{C5b}$$

The term associated with $\mathcal{E}_0(\mathbf{B}_0)$ corresponds to a diagonal matrix and represents a Lamb shift due to the interaction with the environment.

To proceed, we use Eqs. (C2) and (D8) to evaluate $\overline{\mathcal{G}}_S$ and $\overline{\mathcal{G}}_{AS}$. As usual, we ignore the self-field terms. Since $\mathbf{C}^{\text{int}}$ is a symmetric tensor, it is clear that $\overline{\mathcal{G}}_l = \frac{1}{\varepsilon_0}\mathbf{C}^{\text{int}} \cdot \boldsymbol{\alpha}_{\text{np},l}(i\xi) \cdot \mathbf{C}^{\text{int}}$ ($l$=S, AS) with $\boldsymbol{\alpha}_{\text{np},l}$ the symmetric and anti-symmetric parts of the polarizability tensor. From Eq. (D8), the anti-symmetric part is given by $\boldsymbol{\alpha}_{\text{np,AS}} = \alpha_0 \frac{3i\varepsilon_g \hat{\mathbf{u}} \times \mathbf{1}}{(\varepsilon_t + 2)^2 - \varepsilon_g^2}$ with $\alpha_0 = 4\pi R^3$. The symmetric part is determined by the remaining terms in Eq. (D8). Two of the terms (the first and last) are independent of $\mathbf{B}_0$ and hence give rise to a Lamb shift that is independent of the atomic state, with no relevant physical consequences in our analysis. Hence, ignoring this Lamb shift, we can use the replacement $\boldsymbol{\alpha}_{\text{np,S}} \to -\alpha_0 \frac{3(\varepsilon_t + 2)\mathbf{1}_t}{(\varepsilon_t + 2)^2 - \varepsilon_g^2}$ in the evaluation of $\mathcal{E}_0(\mathbf{B}_0)$. The previous considerations show that the matrix elements of the effective Hamiltonian are:

$$h_{m,n} = \delta_{mn}\mathcal{E}_0(\mathbf{B}_0) + \frac{3\alpha_0}{\pi\varepsilon_0}\text{tr}\left\{\mathbf{Q}^{(m,n)}\cdot\mathbf{C}^{\text{int}}\cdot(\hat{\mathbf{u}}\times\mathbf{1})\cdot\mathbf{C}^{\text{int}}\right\}\int_{0^+}^{\infty}d\xi\frac{\varepsilon_g}{(\varepsilon_t+2)^2-\varepsilon_g^2}\frac{\xi}{\omega_a^2+\xi^2}, \quad \text{(C6a)}$$

$$\mathcal{E}_0(\mathbf{B}_0) = \frac{3\alpha_0}{\pi\varepsilon_0}\text{tr}\left\{\mathbf{Q}^{(m,m)}\cdot\mathbf{C}^{\text{int}}\cdot\mathbf{1}_t\cdot\mathbf{C}^{\text{int}}\right\}\int_{0^+}^{\infty}d\xi\frac{(\varepsilon_t+2)}{(\varepsilon_t+2)^2-\varepsilon_g^2}\frac{\omega_a}{\omega_a^2+\xi^2}. \quad \text{(C6b)}$$

The permittivity elements ($\varepsilon_t,\varepsilon_g$) are evaluated with $\omega = i\xi$.

Next, we use $\mathbf{C}^{\text{int}} = \frac{1}{4\pi d^3}(3\hat{\mathbf{z}}\otimes\hat{\mathbf{z}}-\mathbf{1})$ to obtain the following results:

$$\mathbf{C}^{\text{int}}\cdot(\hat{\mathbf{u}}\times\mathbf{1})\cdot\mathbf{C}^{\text{int}} = \frac{1}{16\pi^2 d^6}\mathbf{w}\times\mathbf{1}, \quad \text{with} \quad \mathbf{w} = \hat{\mathbf{u}} + 3(\hat{\mathbf{u}}\times\hat{\mathbf{z}})\times\hat{\mathbf{z}}. \quad \text{(C7a)}$$

$$\mathbf{C}^{\text{int}}\cdot\mathbf{1}_t\cdot\mathbf{C}^{\text{int}} = \frac{1}{16\pi^2 d^6}\left(\mathbf{1}-\hat{\mathbf{u}}\otimes\hat{\mathbf{u}} + (3-9(\hat{\mathbf{u}}\cdot\hat{\mathbf{z}})^2)\hat{\mathbf{z}}\otimes\hat{\mathbf{z}} + 3\hat{\mathbf{u}}\cdot\hat{\mathbf{z}}(\hat{\mathbf{u}}\otimes\hat{\mathbf{z}}+\hat{\mathbf{z}}\otimes\hat{\mathbf{u}})\right). \quad \text{(C7b)}$$

It is recalled that $\hat{\mathbf{u}}$ is the direction of the magnetic field created by the spin magnetic moment of the atom $\mathbf{B}_0$. Evidently, $\mathbf{B}_0$ is rather weak and thereby it is justified to consider its effects perturbatively. From the previous formulas, one can see that $\mathcal{E}_0$ only depends on $|\mathbf{B}_0|^2$, whereas the contribution from the anti-symmetric part of the Green's function (integral in Eq. (C6a)) is proportional to $|\mathbf{B}_0|$. Hence, to leading order one can safely drop the term $\mathcal{E}_0$. With such an approximation, we finally obtain:

$$h_{m,n} \approx -iA\frac{\alpha_0}{16\varepsilon_0\pi^3 d^6}\text{tr}\left\{\mathbf{Q}^{(m,n)}\cdot\left(\frac{\boldsymbol{\omega}_{\text{ef}}}{\omega_p}\times\mathbf{1}\right)\right\}. \quad \text{(C8)}$$

where $\boldsymbol{\omega}_{\text{ef}} = \boldsymbol{\omega}_0 + 3(\boldsymbol{\omega}_0\times\hat{\mathbf{z}})\times\hat{\mathbf{z}}$, $\boldsymbol{\omega}_0 = \omega_0\hat{\mathbf{u}} = \frac{-q}{m^*}\mathbf{B}_0$ is the oriented cyclotron frequency, and $A$ is the dimensionless real-valued parameter

$$A = \int_{0^+}^{\infty}d\xi\frac{i\varepsilon_g/\omega_0}{(\varepsilon_t+2)^2-\varepsilon_g^2}\frac{3\xi\omega_p}{\omega_a^2+\xi^2}. \quad \text{(C9)}$$

Taking into account that $\Gamma \gg \omega_0$ for realistic nanoparticles, one sees from Eq. (6) that the integral can be evaluated using $\varepsilon_t \approx 1 + \frac{\omega_p^2}{\xi(\xi+\Gamma)}$ and $\frac{i\varepsilon_g}{\omega_0} \approx \frac{1}{\xi} \frac{\omega_p^2}{(\xi+\Gamma)^2}$. Furthermore, the term $\varepsilon_g^2$ in the denominator of the integral is negligible. These considerations show that to leading order the parameter $A$ is a constant (i.e., it is independent of the atomic state):

$$A \approx \int_0^\infty d\xi \frac{1}{\left(3\xi(\xi+\Gamma)+\omega_p^2\right)^2} \frac{3\xi^2 \omega_p^3}{\omega_a^2 + \xi^2}. \tag{C10}$$

## Appendix D: The polarizability of the nanoparticle

In this Appendix, we calculate the polarizability of the spherical nanoparticle using a quasi-static approximation (see also Ref. [59]). Specifically, neglecting the effects of time retardation, one may assume that $\mathbf{E} \approx -\nabla\phi$ with $\phi$ an electric potential that satisfies $\nabla \cdot \left(\overline{\varepsilon} \cdot \nabla\phi\right) = 0$. In order to find the electric polarizability, it is supposed that the nanoparticle is illuminated with a constant electric field ($\mathbf{E}^{inc}$) described by the electric potential $-\mathbf{E}^{inc} \cdot \mathbf{r}$. Then, for a nano-particle standing in air the total electric potential may be written as:

$$\phi^{out} = -\mathbf{E}^{inc} \cdot \mathbf{r} + \frac{\mathbf{p}_1 \cdot \hat{\mathbf{r}}}{4\pi\varepsilon_0 r^2}, \qquad r > R \tag{D1a}$$

$$\phi^{int} = -\mathbf{E}_1 \cdot \mathbf{r}, \qquad r < R \tag{D1b}$$

The field outside the nanoparticle is the superposition of the incident field of the field created by an (unknown) electric dipole ($\mathbf{p}_1$). The field inside the nanoparticle ($\mathbf{E}_1$) is a constant. Note that both $\phi^{int}$ and $\phi^{out}$ satisfy $\nabla \cdot (\bar{\varepsilon} \cdot \nabla \phi) = 0$ in the relevant region of space. In order to determine the unknowns $\mathbf{p}_1$ and $\mathbf{E}_1$ one needs to enforce suitable boundary conditions at $r = R$.

The continuity of the electric potential is ensured by:

$$\mathbf{E}^{inc} R - \frac{\mathbf{p}_1}{4\pi\varepsilon_0 R^2} = \mathbf{E}_1 R \tag{D2}$$

The continuity of the electric displacement vector is guaranteed by the condition:

$$\varepsilon_0 \left( \mathbf{E}^{inc} + \frac{2\mathbf{p}_1}{4\pi\varepsilon_0 R^3} \right) = \bar{\varepsilon} \cdot \mathbf{E}_1 \tag{D3}$$

Combining the two boundary conditions, one finds that:

$$\varepsilon_0 \left( \mathbf{E}^{inc} + \frac{2\mathbf{p}_1}{\varepsilon_0 \alpha_0} \right) = \bar{\varepsilon} \cdot \left( \mathbf{E}^{inc} - \frac{\mathbf{p}_1}{\varepsilon_0 \alpha_0} \right) \tag{D4}$$

with $\alpha_0 = 4\pi R^3$. The nanoparticle electric polarizability $\boldsymbol{\alpha}_{np}$ is defined in such a way $\mathbf{p}_1 = \varepsilon_0 \boldsymbol{\alpha}_{np} \cdot \mathbf{E}^{inc}$. Clearly, the polarizability satisfies:

$$\mathbf{1} + \frac{2\boldsymbol{\alpha}_{np}}{\alpha_0} = \frac{\bar{\varepsilon}}{\varepsilon_0} \cdot \left( \mathbf{1} - \frac{\boldsymbol{\alpha}_{np}}{\alpha_0} \right) \tag{D5}$$

Solving for $\boldsymbol{\alpha}_{np}$ one finally concludes that:

$$\boldsymbol{\alpha}_{np} = \alpha_0 \left( \frac{\bar{\varepsilon}}{\varepsilon_0} + 2\mathbf{1}_{3\times 3} \right)^{-1} \cdot \left( \frac{\bar{\varepsilon}}{\varepsilon_0} - \mathbf{1}_{3\times 3} \right). \tag{D6}$$

The electric polarizability may also be written as:

$$\boldsymbol{\alpha}_{np} = \alpha_0 \left[ \mathbf{1}_{3\times 3} - 3 \left( \frac{\bar{\varepsilon}}{\varepsilon_0} + 2\mathbf{1}_{3\times 3} \right)^{-1} \right]. \tag{D7}$$

It should be mentioned that the quasi-static polarizability does not account for the effect of radiation loss, which is typically negligible as compared to the material loss.

For the particular case of a gyroelectric permittivity response as in Eq. (5) the polarizability can be explicitly written as:

$$\boldsymbol{\alpha}_{np} = \alpha_0 \left[ \mathbf{1}_{3\times 3} - \frac{3}{(\varepsilon_t + 2)^2 - \varepsilon_g^2} \left[ (\varepsilon_t + 2)\mathbf{1}_t - i\varepsilon_g \hat{\mathbf{u}} \times \mathbf{1} \right] - \frac{3}{\varepsilon_a + 2} \hat{\mathbf{u}} \otimes \hat{\mathbf{u}} \right], \quad (D8)$$

with $\alpha_0 = 4\pi R^3$.

## Appendix E: Lindblad master equation with radiation loss

In this Appendix, we generalize the quantum master equation (B4) in order to include perturbatively the effects of the radiation loss due to the precession of the atomic spin magnetic moment.

*A. Master equation without radiation loss*

To begin with, it is useful to write explicitly Eq. (B4) in terms of the density matrix elements $\hat{\rho}_S(t) = \sum_{i,j} \rho_{ij} |g_i\rangle\langle g_j|$ and of the effective Hamiltonian elements (Eq. (19)). Taking into account that $h_{22} = -h_{11}$ one finds that:

$$\partial_t \begin{pmatrix} \rho_{11} & \rho_{12} \\ \rho_{21} & \rho_{22} \end{pmatrix} = -i\frac{1}{\hbar} \begin{pmatrix} h_{12}\rho_{21} - h_{21}\rho_{12} & 2h_{11}\rho_{12} + h_{12}(\rho_{22} - \rho_{11}) \\ -2h_{11}\rho_{21} - h_{21}(\rho_{22} - \rho_{11}) & -(h_{12}\rho_{21} - h_{21}\rho_{12}) \end{pmatrix}. \quad (E1)$$

The elements $h_{ij}$ are defined as in Eq. (19) with the replacements $c_i c_j^* \to \rho_{ij}$, so that $h_{11} = \pm \mathcal{E}_d (\rho_{11} - \rho_{22})$ and $h_{12} = h_{21}^* = \pm \mathcal{E}_{cr} \rho_{ba}$ with $(a,b) = (1,2)$ for the $\mathbf{H}_{ef,+}$ system and $(a,b) = (2,1)$ for the $\mathbf{H}_{ef,-}$ system. The time evolution of both pure and mixed atomic states can be determined using the above equation.

## B. Radiation loss

Next, we take into account the effects of radiation loss perturbatively. A time crystal state originates a time-varying magnetic dipole moment, and hence emits radiation to the far-field. It should be noted that in our model the electric dipole moment vanishes in the ground subspace and hence cannot emit radiation.

The derivation of the master equation up to the end of subsection C of Appendix A is completely general. In particular, Eq. (A11) incorporates the magnetic dipole ($\hat{\mathbf{p}}_m$) effects in the operator $\hat{\mathbf{p}} = (\hat{\mathbf{p}}_e \quad \hat{\mathbf{p}}_m)^T$. We recall that $\hat{\mathbf{p}}(t)$ represents the dipole operator in the interaction picture ($\hat{\mathbf{p}}(t) = \exp\left(+\frac{i}{\hbar}\hat{H}_{at}t\right)\hat{\mathbf{p}}\exp\left(-\frac{i}{\hbar}\hat{H}_{at}t\right)$). Since $(\hat{\mathbf{p}}_m)_i = \mu_0 \frac{\hbar}{2}\frac{q}{m_e}\boldsymbol{\sigma}_i$ with $\boldsymbol{\sigma}_i$ the Pauli matrix, it follows that $\hat{\mathbf{p}}_m(t) = \hat{\mathbf{p}}_m$ in the ground subspace. In other words, the operator $\hat{\mathbf{p}}_m(t)$ remains independent of time in the interaction picture. It is possible to circumvent this deficiency, by incorporating the effective Hamiltonian into the atom Hamiltonian. With this approximation, we find:

$$\hat{\mathbf{p}}_m(t) \approx \exp\left(+\frac{i}{\hbar}\hat{H}_{ef}t\right)\hat{\mathbf{p}}_m\exp\left(-\frac{i}{\hbar}\hat{H}_{ef}t\right). \tag{E2}$$

For simplicity, we restrict our analysis to the case wherein $\hat{H}_{ef}$ is approximately diagonal ($h_{12} \approx 0$) and $h_{22} = -h_{11}$ ($|\gamma_c| \ll |\gamma_d|$). Then, straightforward calculations show that:

$$\hat{\mathbf{p}}_m(t) = \hat{\mathbf{p}}_{m0} + \hat{\mathbf{p}}_m^- e^{-i\omega_m t} + \hat{\mathbf{p}}_m^+ e^{+i\omega_m t} \tag{E3a}$$

with $\hat{\mathbf{p}}_{m0}$ some operator independent of time, $\hat{\mathbf{p}}_m^- = \boldsymbol{\gamma}_m |g_2\rangle\langle g_1|$, $\hat{\mathbf{p}}_m^+ = (\hat{\mathbf{p}}_m^-)^\dagger$, and

$$\boldsymbol{\gamma}_{\mathrm{m}} = \mu_0 \frac{\hbar}{2} \frac{q}{m_e} \begin{pmatrix} 1 \\ +i \\ 0 \end{pmatrix}, \qquad \omega_{\mathrm{m}} = \frac{2h_{11}}{\hbar}. \qquad (E3b)$$

Note that $\omega_{\mathrm{m}}$ is precisely the spin precession frequency when $|\gamma_{\mathrm{c}}| \ll |\gamma_{\mathrm{d}}|$ [Eq. (22)]. As in Appendix A, we include a small positive imaginary part in $\omega_{\mathrm{m}}$ to model an adiabatic switching of the interaction term. With this modification, the magnetic dipole operator becomes $\hat{\mathbf{p}}_{\mathrm{m}}(t) \to \hat{\mathbf{p}}_{\mathrm{m}}^{-} e^{-i\omega_{\mathrm{m}} t} + \hat{\mathbf{p}}_{\mathrm{m}}^{+} e^{+i\omega_{\mathrm{m}}^{*} t}$, where we dropped the time independent term $\hat{\mathbf{p}}_{\mathrm{m}0}$ which plays no role in the dynamics. Substituting $\hat{\mathbf{p}}_{\mathrm{m}}(t) \to \hat{\mathbf{p}}_{\mathrm{m}}^{-} e^{-i\omega_{\mathrm{m}} t} + \hat{\mathbf{p}}_{\mathrm{m}}^{+} e^{+i\omega_{\mathrm{m}}^{*} t}$ into Eq. Eq. (A11), the Lindblad operator in the Schrödinger picture gets an additional term ($\mathcal{L}_{\mathrm{m}}$) due to the magnetic dipole radiation such that (compare with Eq. (A15)):

$$\begin{aligned}
\mathcal{L}_{\mathrm{m}} \hat{\rho}_{\mathrm{S}} = &-\frac{i}{\hbar} \hat{\rho}_{\mathrm{S}}(t) \left( \hat{\mathbf{p}}_{\mathrm{m}}^{-} \cdot \left[ \overline{\mathcal{G}}_{\mathrm{mm}}^{-} \right]^{*} \cdot \hat{\mathbf{p}}_{\mathrm{m}} + \hat{\mathbf{p}}_{\mathrm{m}}^{+} \cdot \left[ \overline{\mathcal{G}}_{\mathrm{mm}}^{+} \right]^{\dagger} \cdot \hat{\mathbf{p}}_{\mathrm{m}} \right) \\
&+ \frac{i}{\hbar} \left( \hat{\mathbf{p}}_{\mathrm{m}} \cdot \overline{\mathcal{G}}_{\mathrm{mm}}^{+} \cdot \hat{\mathbf{p}}_{\mathrm{m}}^{-} + \hat{\mathbf{p}}_{\mathrm{m}} \cdot \left[ \overline{\mathcal{G}}_{\mathrm{mm}}^{-} \right]^{T} \cdot \hat{\mathbf{p}}_{\mathrm{m}}^{+} \right) \hat{\rho}_{\mathrm{S}}(t) \\
&- \frac{i}{\hbar} \left( \hat{\mathbf{p}}_{\mathrm{m}}^{-} \cdot \left[ \overline{\mathcal{G}}_{\mathrm{mm}}^{+} \right]^{T} \hat{\rho}_{\mathrm{S}}(t) \cdot \hat{\mathbf{p}}_{\mathrm{m}} + \hat{\mathbf{p}}_{\mathrm{m}}^{+} \cdot \overline{\mathcal{G}}_{\mathrm{mm}}^{-} \hat{\rho}_{\mathrm{S}}(t) \cdot \hat{\mathbf{p}}_{\mathrm{m}} \right) \\
&+ \frac{i}{\hbar} \left( \hat{\mathbf{p}}_{\mathrm{m}} \cdot \left[ \overline{\mathcal{G}}_{\mathrm{mm}}^{-} \right]^{\dagger} \hat{\rho}_{\mathrm{S}}(t) \cdot \hat{\mathbf{p}}_{\mathrm{m}}^{-} + \hat{\mathbf{p}}_{\mathrm{m}} \cdot \left[ \overline{\mathcal{G}}_{\mathrm{mm}}^{+} \right]^{*} \hat{\rho}_{\mathrm{S}}(t) \cdot \hat{\mathbf{p}}_{\mathrm{m}}^{+} \right).
\end{aligned} \qquad (E4)$$

Here, $\overline{\mathcal{G}}_{\mathrm{mm}}^{\pm}$ are defined as in Eq. (A16) with the replacement $\omega_{\mathrm{a}} \to \omega_{\mathrm{m}}$. Furthermore, $\overline{\mathcal{G}}_{\mathrm{mm}}^{\pm}$ refer to the magnetic part of the Green's function (right-lower 3×3 sub-block of the 6×6 tensor). Similar to subsection E of Appendix A, $\overline{\mathcal{G}}_{\mathrm{mm}}^{\pm}$ can be expressed in terms of the system Green's function. However, one needs to be careful because here $\omega_{\mathrm{m}} = \frac{2h_{11}}{\hbar}$ can be either be positive or negative, depending on the system state. For a positive $\omega_{\mathrm{m}}$, the results in Eqs. (A20) and (A21) hold true. We can ignore the non-resonant part of the Green's function ($\overline{\mathcal{G}}_{mm}^-$) as it is associated with a Lamb shift. Thus, for $\omega_m > 0$ we can replace $\overline{\mathcal{G}}_{mm}^- \to 0$ and $\overline{\mathcal{G}}_{mm}^+ \to \overline{\mathcal{G}}_{mm}$ in Eq. (E4). Since the nanoparticle does not have a magnetic response, the Green's function can be identified with the free-space Green's function, so that $\operatorname{Im}\{\overline{\mathcal{G}}_{mm}\} = \frac{1}{6\pi\mu_0}\left(\frac{\omega_m}{c}\right)^3 \mathbf{1}_{3\times 3}$. We recall that the imaginary part of the Green's function controls the spontaneous emission decay in free-space. In summary, for $\omega_m > 0$ the radiation from the magnetic dipole precession is modeled by an operator $\mathcal{L}_m$ such that:

$$\mathcal{L}_m \hat{\rho}_S = \frac{1}{6\pi\mu_0 \hbar}\left(\frac{\omega_m}{c}\right)^3 \left[ -\hat{\rho}_S(t) \hat{\mathbf{p}}_m^+ \cdot \hat{\mathbf{p}}_m^- - \hat{\mathbf{p}}_m^+ \cdot \hat{\mathbf{p}}_m^- \hat{\rho}_S(t) + 2 \hat{\mathbf{p}}_m^- \cdot \hat{\rho}_S(t) \cdot \hat{\mathbf{p}}_m^+ \right], \quad \omega_m > 0 \quad (E5)$$

We took into account that $\hat{\mathbf{p}}_m^+ \cdot \hat{\mathbf{p}}_m = \hat{\mathbf{p}}_m^+ \cdot \hat{\mathbf{p}}_m^- = \hat{\mathbf{p}}_m \cdot \hat{\mathbf{p}}_m^-$ and that $\hat{\mathbf{p}}_m^- \cdot \hat{\rho}_S(t) \cdot \hat{\mathbf{p}}_m = \hat{\mathbf{p}}_m^- \cdot \hat{\rho}_S(t) \cdot \hat{\mathbf{p}}_m^+$ and $\hat{\mathbf{p}}_m \cdot \hat{\rho}_S(t) \cdot \hat{\mathbf{p}}_m^+ = \hat{\mathbf{p}}_m^- \cdot \hat{\rho}_S(t) \cdot \hat{\mathbf{p}}_m^+$. Clearly, the formula for $\omega_m < 0$ can be obtained from the previous one with the replacements $\omega_m \to -\omega_m$, $\hat{\mathbf{p}}_m^- \to \hat{\mathbf{p}}_m^+$, and $\hat{\mathbf{p}}_m^+ \to \hat{\mathbf{p}}_m^-$. So, the operator is generally defined as:

$$\mathcal{L}_m \hat{\rho}_S = -\frac{\Gamma_{sp}}{2} \begin{cases} \hat{\rho}_S(t)|g_1\rangle\langle g_1| + |g_1\rangle\langle g_1|\hat{\rho}_S(t) - 2|g_2\rangle\langle g_1|\hat{\rho}_S(t)|g_1\rangle\langle g_2|, & \omega_m > 0 \\ \hat{\rho}_S(t)|g_2\rangle\langle g_2| + |g_2\rangle\langle g_2|\hat{\rho}_S(t) - 2|g_1\rangle\langle g_2|\hat{\rho}_S(t)|g_2\rangle\langle g_1|, & \omega_m < 0 \end{cases} \quad (E6)$$

where $\Gamma_{sp} = 2\frac{|\gamma_m|^2}{6\pi\mu_0 \hbar}\left(\frac{|\omega_m|}{c}\right)^3$ is the spontaneous emission rate due to the magnetic spin precession in free-space. Using $|\gamma_m|^2 = 2\left(\mu_0 \frac{\hbar}{2} \frac{q}{m_e}\right)^2$ the decay rate $\Gamma_{sp}$ can also be written as

$$\Gamma_{\text{sp}} = \alpha_{\text{e,fs}} \frac{2}{3}|\omega_{\text{m}}|\left(\frac{\hbar|\omega_{\text{m}}|}{m_e c^2}\right)^2 = g_{\text{sp}} \frac{2\mathcal{E}_{\text{d}}}{\hbar}|\rho_{11} - \rho_{22}|^3, \qquad (E7)$$

where $\alpha_{\text{e,fs}}$ is the electron fine structure constant and $g_{\text{sp}} = \alpha_{\text{e,fs}} \frac{8}{3}\left(\frac{\mathcal{E}_{\text{d}}}{m_e c^2}\right)^2$ is a dimensionless factor. To conclude, we note that $\mathcal{L}_{\text{m}}\hat{\rho}_{\text{S}}$ is represented by the following matrix in the spin up and spin down basis:

$$\mathcal{L}_{\text{m}}\hat{\rho}_{\text{S}} \to -\Gamma_{\text{sp}} \begin{pmatrix} \tilde{\rho} & \rho_{12}/2 \\ \rho_{21}/2 & -\tilde{\rho} \end{pmatrix} \quad \text{with} \quad \tilde{\rho} = \begin{cases} \rho_{11}, & \omega_{\text{m}} > 0 \\ -\rho_{22}, & \omega_{\text{m}} < 0 \end{cases}. \qquad (E8)$$

The full master equation is $\partial_t \hat{\rho}_{\text{S}} = -i\frac{1}{\hbar}\left[\hat{H}_{\text{ef}}, \hat{\rho}_{\text{S}}\right] + \mathcal{L}_{\text{m}}\hat{\rho}_{\text{S}}$ and can be written explicitly by adding to the right-hand side of Eq. (E1) the matrix $\mathcal{L}_{\text{m}}\hat{\rho}_{\text{S}}$ defined in the above equation.